\documentclass[10pt,twocolumn,letterpaper]{article}

\usepackage{cvpr}
\usepackage{times}
\usepackage{epsfig}
\usepackage{graphicx}
\usepackage{amsmath}
\usepackage{amssymb}
\usepackage{booktabs}
\usepackage{subfig}
\usepackage[linesnumbered,ruled]{algorithm2e}
\usepackage{soul}
\usepackage{ulem}
\usepackage{color}
\usepackage{csquotes}
\usepackage{comment}

\definecolor{CQColor}{rgb}{0.0,0.0,1.0} 

\usepackage[breaklinks=true,bookmarks=false]{hyperref}

\cvprfinalcopy 

\def\cvprPaperID{4587} 

\begin{document}

\title{Flexible Attributed Network Embedding}
\author{Enya Shen\\
Tsinghua University, China\\
{\tt\small shenenya@tsinghua.edu.cn}
\and
Zhidong Cao\\
University of California, San Diego, US\\
{\tt\small zhc178@ucsd.edu}
\and
Changqing Zou\\
University of Maryland, College Park, US\\
{\tt\small aaronzou1125@gmail.com}
\and
Jianmin Wang\\
Tsinghua University, China\\
{\tt\small jimwang@tsinghua.edu.cn}
}
\maketitle

\begin{abstract}
Network embedding aims to find a way to encode network by learning an embedding vector for each node in the network. The network often has property information which is highly informative with respect to the node's position and role in the network. Most network embedding methods fail to utilize this information during network representation learning. In this paper, we propose a novel framework, FANE, to integrate structure and property information in the network embedding process. In FANE, we design a network to unify heterogeneity of the two information sources, and define a new random walking strategy to leverage property information and make the two information compensate. FANE is conceptually simple and empirically powerful. It improves over the state-of-the-art methods on Cora dataset classification task by over 5\%, more than 10\% on WebKB dataset classification task. Experiments also show that the results improve more than the state-of-the-art methods as increasing training size. Moreover, qualitative visualization show that our framework is helpful in network property information exploration. In all, we present a new way for efficiently learning state-of-the-art task-independent representations in complex attributed networks. The source code and datasets of this paper can be obtained from \url{https://github.com/GraphWorld/FANE}.
\end{abstract}

\section{Introduction} \label{sec:introduction}
Network embedding is an important and ubiquitous research problem with applications ranging from drug design to commodities or friendship recommendations~\cite{RN20104, RN20123, RN20012, RN20015}. For most practical networks in the form of graphs, nodes have more than one attributes that greatly determine their roles in the system. For example, individuals in a social network have various properties, such as sexuality, educational background, and partisanship. Moreover, the social science \cite{RN20248, RN20249} has shown that attributes of nodes can reflect and affect their community structures \cite{GaoH18b, Huang-etal17Accelerated, RN20250}.

Current network embedding studies include structure-preserving methods and property-preserving methods \cite{Du2018}.
After decades of research, many structure-preserving approaches, such as DeepWalk \cite{RN20110}, node2vec \cite{RN20017} and struc2vec \cite{RN20124}, have been proposed to learn network features based on structure information. However, these methods only consider network structures, failing to take advantages of node attributes during encoding \cite{RN20104}.

Recently, few property-preserving methods are proposed. These methods could be further categorized into matrix factorization and deep learning based methods. Matrix factorization based network embedding represent network property in the form of a matrix and factorize this matrix to obtain node embedding \cite{RN20123}, such as TADW \cite{RN20134} and HSCA \cite{RN20136}. These methods is time and space consuming \cite{RN20012}.
Deep learning based network embedding, such as SNE \cite{RN20129}, DANE \cite{GaoH18b} and DVNE \cite{Zhu2018}, get inspiration from existing neural network models and (or) design new model to learn network features. These methods get high accuracy result at the cost of high training time requirement.

In response, we propose FANE, a scalable and flexible attributed network embedding framework to integrate both structure and property information to learn features. Briefly, we design a network to unify heterogeneity of the structure and property information sources, and define a new random walking strategy to leverage property information and make them compensate and flexible. Overall, our paper makes the following contributions:

\begin{itemize}
\item We propose FANE, an efficient and flexible framework that integrates network attribute and structure for feature learning in networks.
\item We analyze and verify that FANE can learn features as state of the art structure-preserving and property-preserving methods.
\item We extend our method onto attribute space. Relationships between attributes can also be explored under our framework.
\item We evaluate our framework on multi-label classification task and conduct visual analysis on several real-world datasets.
\end{itemize}
\section{Related works} \label{sec:related-works}
One key problem in network embedding is what to preserve in learning. Here we discuss related works based on what they aim to preserve.

\textbf{Structure-preserving network embedding.}
DeepWalk \cite{RN20110} generalizes language modeling SkipGram \cite{RN20126} for network embedding, which uses random walks to learn latent representations by treating walks as the equivalent of sentences. Instead of exploiting random walks to capture network structure, LINE \cite{RN20125} learns vertex representations by explicitly modeling the first-order and second-order proximity. In addition, struc2vec \cite{RN20124} first encodes the vertex structural role similarity into a multilayer network, where the weights of edges at each layer are determined by the structural role difference at the corresponding scale. Moreover, node2vec \cite{RN20017} presents a random walking method to interpolates between Breadth-first Sampling and Depth-first Sampling. Inspired by those methods, we have designed a new random walking strategy to leverage property information and make them compensate in the embedding process.

\textbf{Property-preserving network embedding.}
TADW \cite{RN20134} extends DeepWalk \cite{RN20110} to get a vertex-context matrix. HSCA \cite{RN20136} integrates homophily, structural context, and vertex content to learn effective network representations. However, calculate factorization on large real-world network matrix with millions of rows and columns is expensive and unscalable. SNE \cite{RN20129} includes two similar deep neural network models to deal with structure and attribute information separately in embedding layer, the result then processed by same hidden layer to learn features. Differently, DANE \cite{GaoH18b} use two separate deep neural network models for structure and attribute information, and use joint distribution to optimize the result. DVNE \cite{Zhu2018} learns a Gaussian distribution in the Wasserstein space as the latent representation of each node. These methods get high accuracy result at the cost of high training time requirement.

\textbf{Attributed network clustering.}
Network clustering aims to divide a given set of objects into groups of similar objects. Lots of related algorithms have been proposed, such as Minimum Cut Algorithm \cite{RN20141}, Multi-way network Partition \cite{RN20142}, k-medoid and k-means algorithm \cite{RN20143}, Spectral Clustering method \cite{RN20140}. SA-cluster \cite{RN20115} and its extended version \cite {RN19978, RN20108} presented a way to class attributed network by extracting attributes as separate node. Inspired by these works, we construct a new network to integrate structure and attribute for network embedding.
\section{FANE} \label{sec:fane}
In this section, we first give the problem definition of attributed network embedding, and then discuss our solutions for the key challenges.

\subsection{Problem definition}
An attributed network is formally denoted as $G = (V, E, W, \Lambda)$, where $V$ is the set of vertices, $E$ is the set of edges, $W$ is the weights of edges, and $\Lambda = \{\alpha_1, ..., \alpha_m\}$ is the set of attributes associated with vertices in $V$ for describing vertex properties. Each vertex $v_i \in V$ is associated with an attribute vector $[\alpha_1(v_i), ..., \alpha_m(v_i)]$ where $\alpha_j(v_i)$ is the attribute value of vertex $v_i$ on attribute $\alpha_j$.

Our goal is to learn a mapping function from network to feature representations $f: G \rightarrow \mathbb{R}^d$. Here $d$ is the dimension of the feature representations. $f$ should support:
\begin{enumerate}
\item[T1:] \textbf{Scalability}. As a network compression approach, The network embedding inherently should be scalable to deal with large scale of network. Meanwhile, the size of network nodes ranges from tens to millions, even billions in practise \cite{RN20012}.
\item[T2:] \textbf{Integration}. Network structures and properties are the fundamental factors that need to be considered in network embedding. However, preserving these properties in a network embedding space is challenging due to the disparity and inhomogeneity between the network space and the embedding vector space \cite{Cui2018}.
\item[T3:] \textbf{Adaption}. Various domain, data and applications require different network embedding methods, such as structure-preserving, property-preserving or both. Is there a way to provide an framework which is flexible enough to learn features as needed?
\end{enumerate}

We proceed by extending the scalable skip-gram architecture (T1) to attributed networks \cite{RN20126, RN20110, RN20017}. Given node $u$ and its context nodes $N_s(u) \subseteq V$, the representation for $u$ is learned by maximizing the conditional probability (objective function):
\begin{equation}
  \prod_{u \in V} P(N_s(u)|f(u))
\end{equation}

The key challenge is how to define the context nodes $N_s(u)$ for \textit{attributed} nodes, so we are able to integrate property information effectively (T2). We deal with this by construct a new network as discussed below.

\subsection{Network construction} \label{construct}
As stated, the goal of network construction here is to effectively integrate structure and property information. But how to get there. Before that, let us go back to the original network and take another way of looking at the problem. In network, different nodes are connected with edges, which in fact reflect the connection between these nodes in property space (structure information). Meanwhile, if two nodes have same attribute, we could say that these nodes have connection in attribute space (property information). So attributes play as edge in attribute space, like \textit{actual} edges in network. In order to integrate attribute information in network embedding. We try to concrete the connection in attribute space by appending special \textit{virtual} edges to represent it. So the heterogeneous edges in the result network could be used to reflect connections between nodes in both structure and property space.

\begin{figure}
\centering
\includegraphics[width=0.45\textwidth]{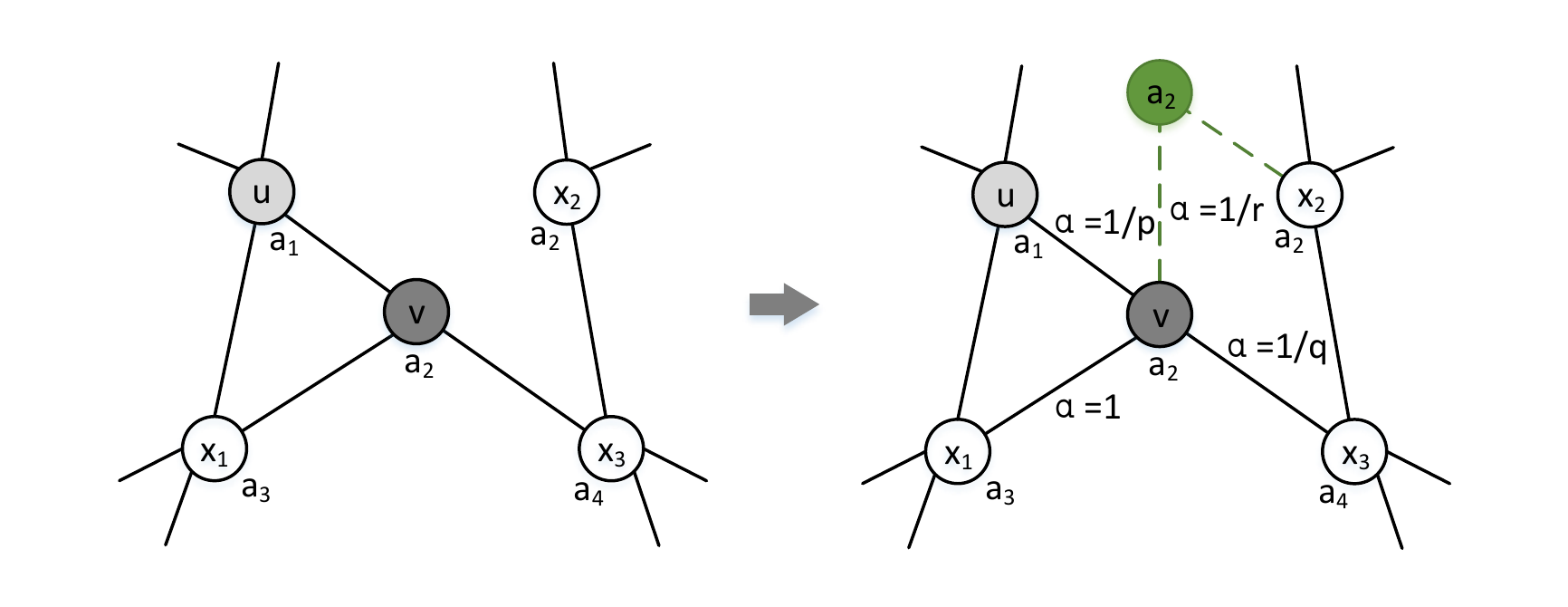}
\caption{Illustration of FANE. The FANE network (right) is constructed by mapping node attributes from the original network (left) to \textit{virtual} attribute nodes. Nodes and their corresponding attribute nodes are connected via \textit{virtual} edges. It is convenient to take advantage of the state-of-the-art network embedding methods, such as node2vec \cite{RN20017}, struc2vec \cite{RN20124}, to leverage attribute information in diverse network embedding applications.}
\label{fig:Attri2Vec}
\end{figure}

Consequently, the simplest way is to add an edge between nodes that have same attribute. However, by doing so, we increase the worst case edge size from $O(n_E)$ to $O(n_E+n_V^2*n_\Lambda$). We resolve the worst case edge size problem by introducing \textit{virtual} attribute nodes. As shown in Figure \ref{fig:Attri2Vec}, attribute network is constructed based on attribute information from the raw network by taking attributes $\{\alpha_1, ..., \alpha_m\}$ as special \textit{virtual} nodes $\{v_{\alpha_1}, ..., v_{\alpha_m}\}$: if node $v_i$ has attribute $\alpha_j$, there will exist a \textit{virtual} edge between $v_i$ and $v_{\alpha_j}$. Thus, the worst case edge size of resultant network would be reduced to $O(n_E+n_V*n_\Lambda$), which is less than $O(n_E+n_V^2*n_\Lambda$) generally.

\begin{figure*}
\centering
\subfloat[FANE-sf]{\includegraphics[width=\textwidth]{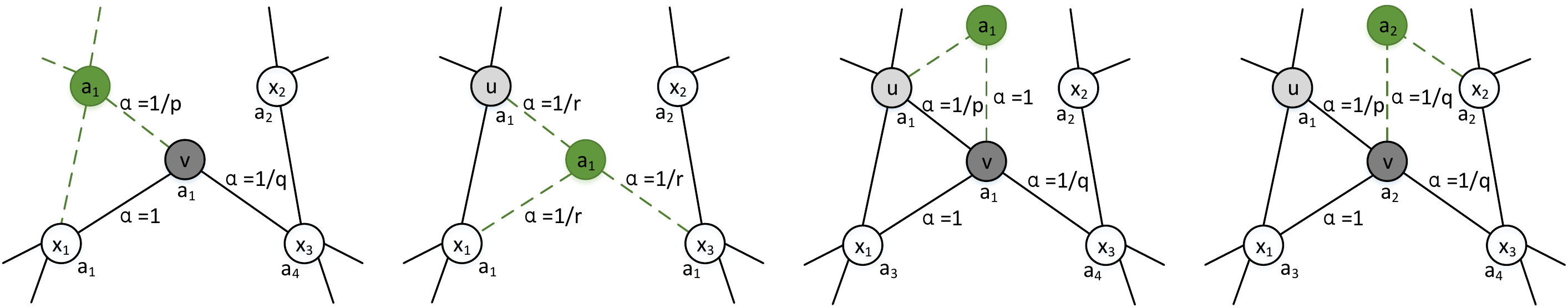}\label{fig:FANE-sf}}\newline
\subfloat[FANE-tf]{\includegraphics[width=\textwidth]{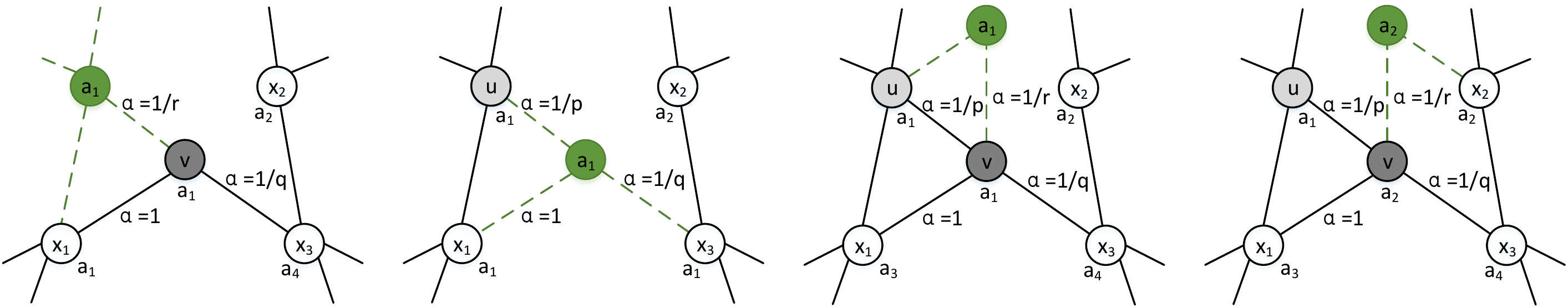}\label{fig:FANE-tf}}\newline
\subfloat[FANE-stf]{\includegraphics[width=\textwidth]{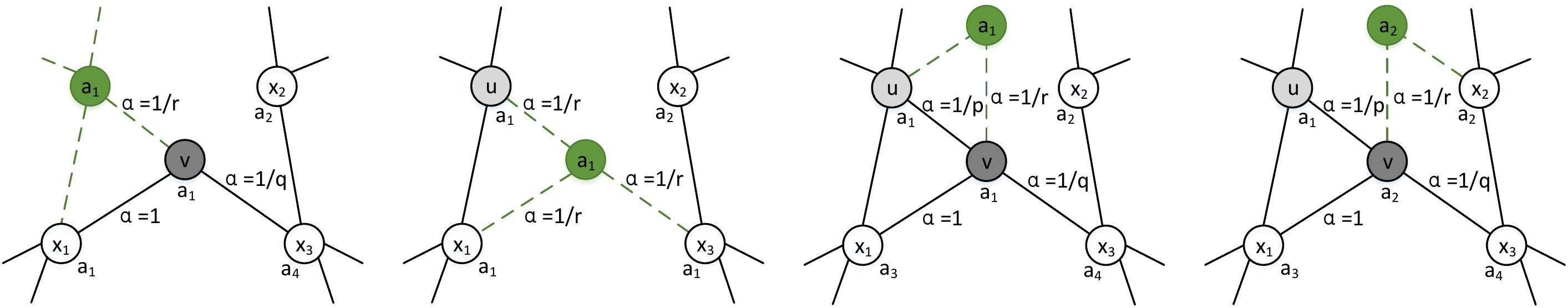}\label{fig:FANE-stf}}
\caption[]{Illustrations of Source-Focused (FANE-sf), Target-Focused (FANE-tf) and Source and Target Focused (FANE-stf) strategies in FANE from top to bottom respectively. The upper-left (green) and the middle nodes (gray) of the network represent the previous and source node separately. The rest are the target nodes. Figures from left to right demonstrate various scenarios in random walking regard to all three strategies.}
\end{figure*}

\begin{figure*}
\centering
\subfloat[FANE-sf]{\includegraphics[width=0.3\textwidth, keepaspectratio]{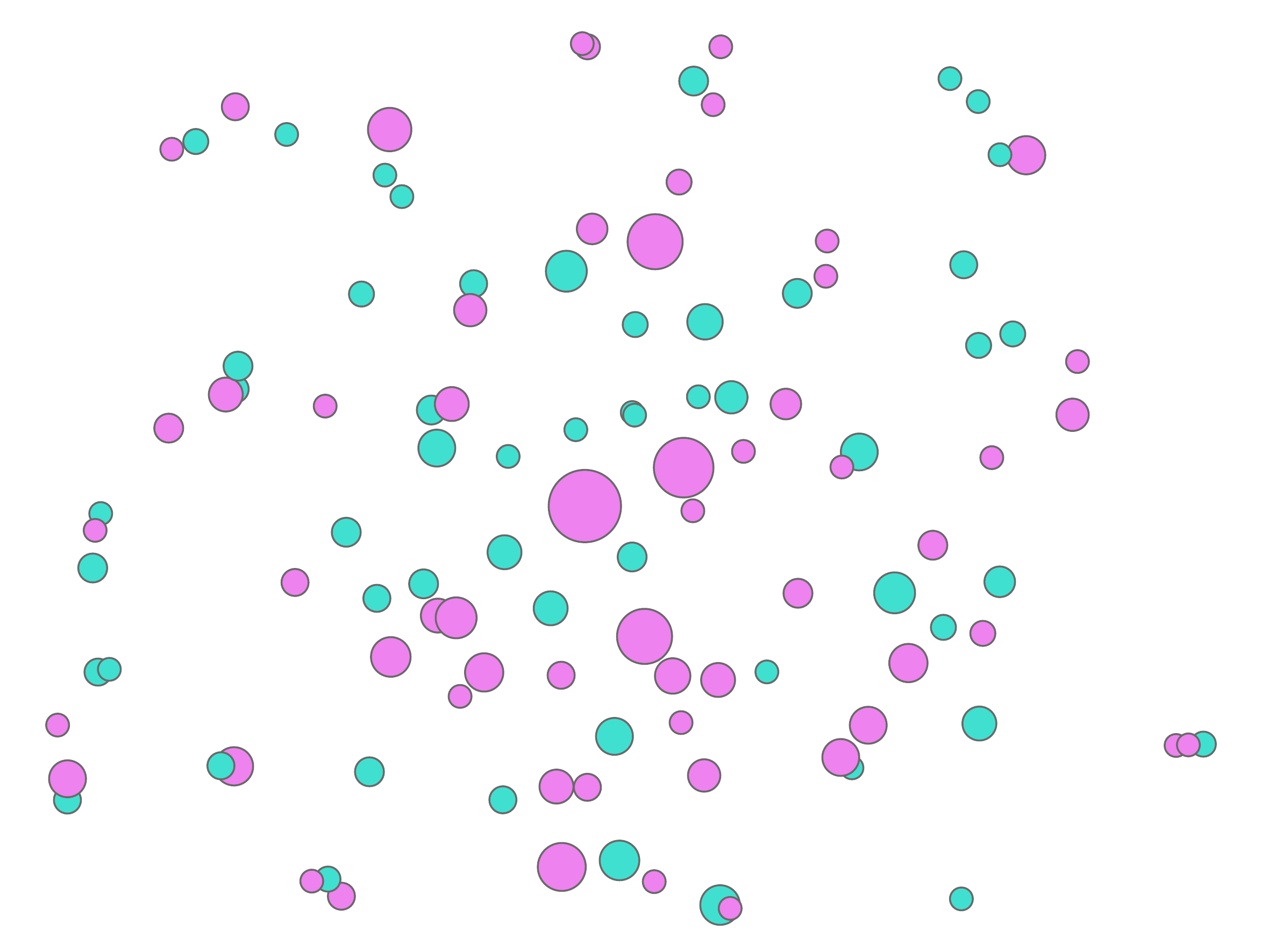}\label{fig:attri2vec-sf}}
\subfloat[FANE-tf]{\includegraphics[width=0.3\textwidth, keepaspectratio]{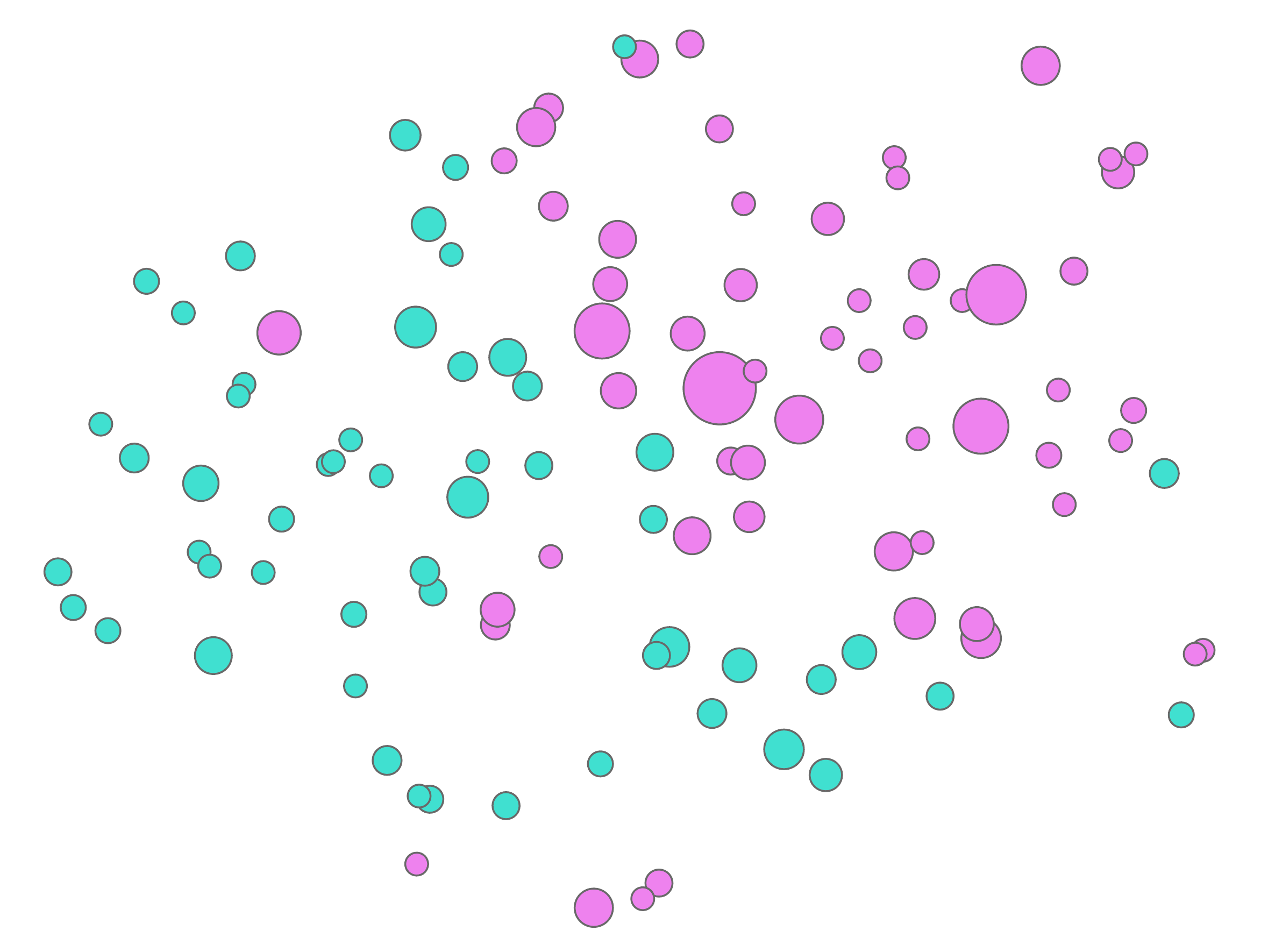}\label{fig:attri2vec-tf}}
\subfloat[FANE-stf]{\includegraphics[width=0.3\textwidth, keepaspectratio]{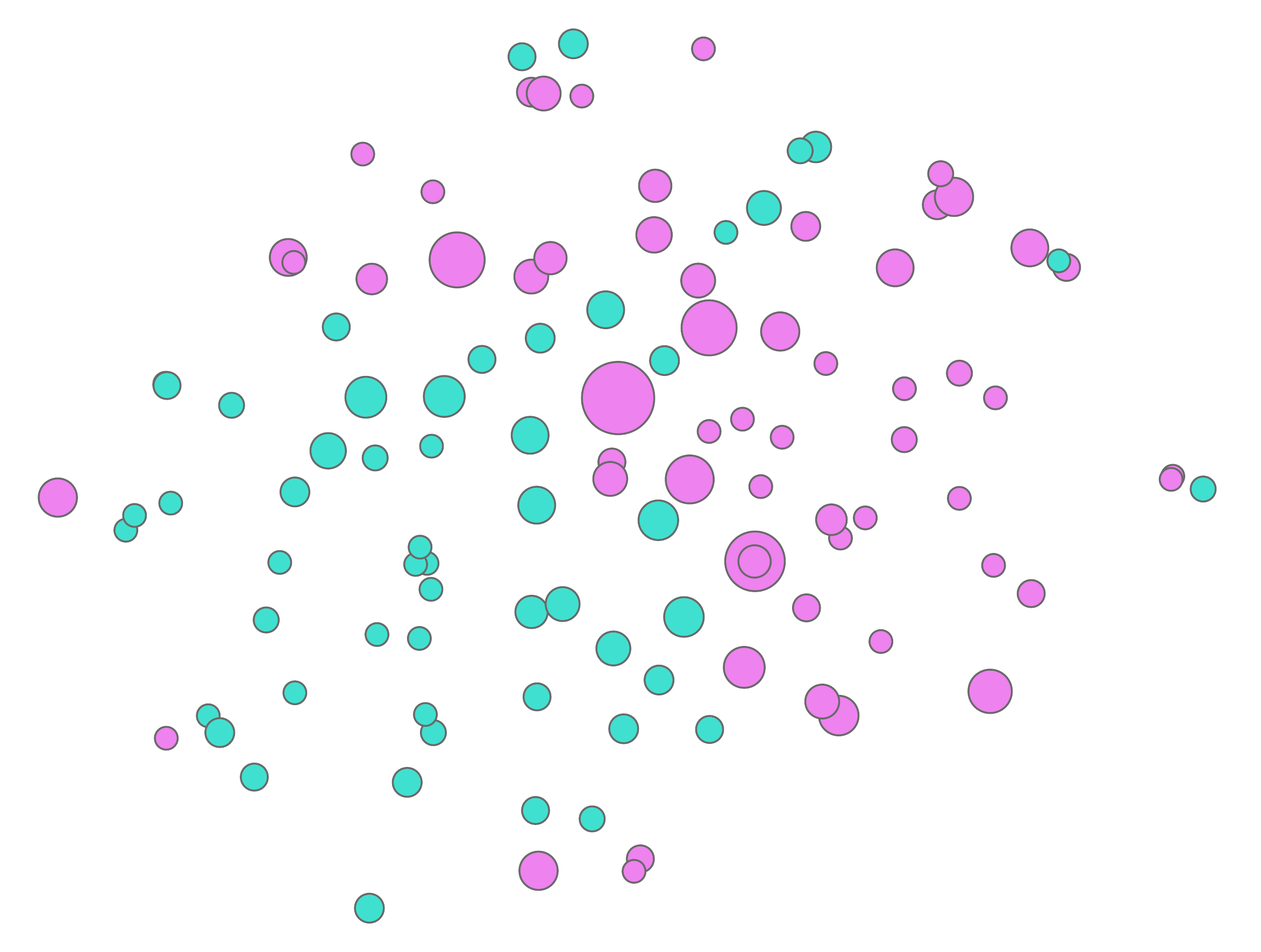}\label{fig:attri2vec-stf}}
\caption[]{Visual analysis of different random walking strategies of FANE with same dimensionality reduction technique t-SNE \cite{RN20145}. Node colors represent different attributes, word classes, of English words \cite{RN20144}. As we can see, part (b-c), which corresponding to the strategy 2 and 3, yield distinct separation between two different attributes.}
\label{fig:attri2vec-f}
\end{figure*}

\begin{figure*}
\centering
\subfloat[$r=2.0$]{\includegraphics[width=0.23\textwidth, keepaspectratio]{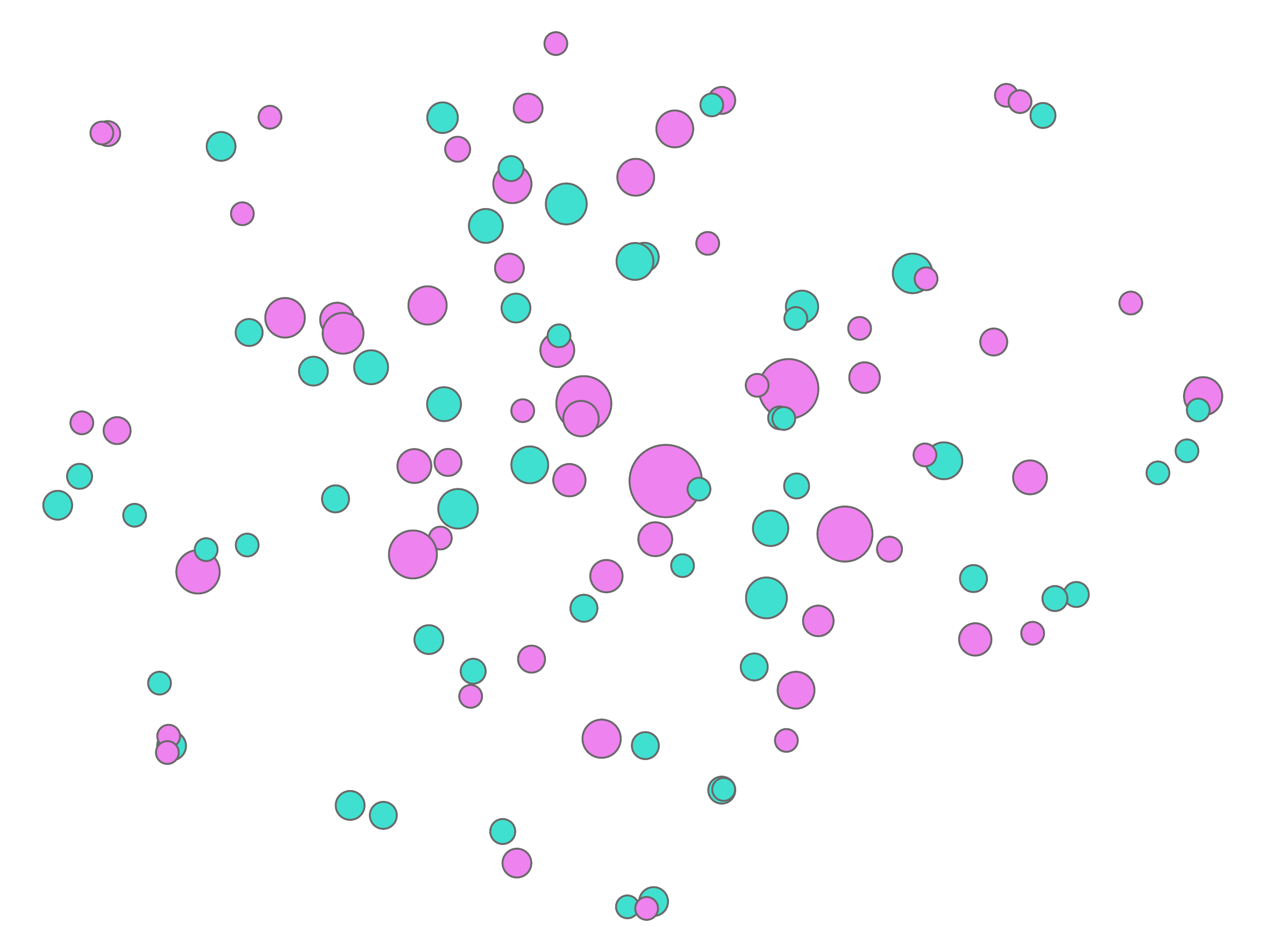}\label{fig:adjnoun-r2_0}}
\subfloat[$r=0.5$]{\includegraphics[width=0.23\textwidth, keepaspectratio]{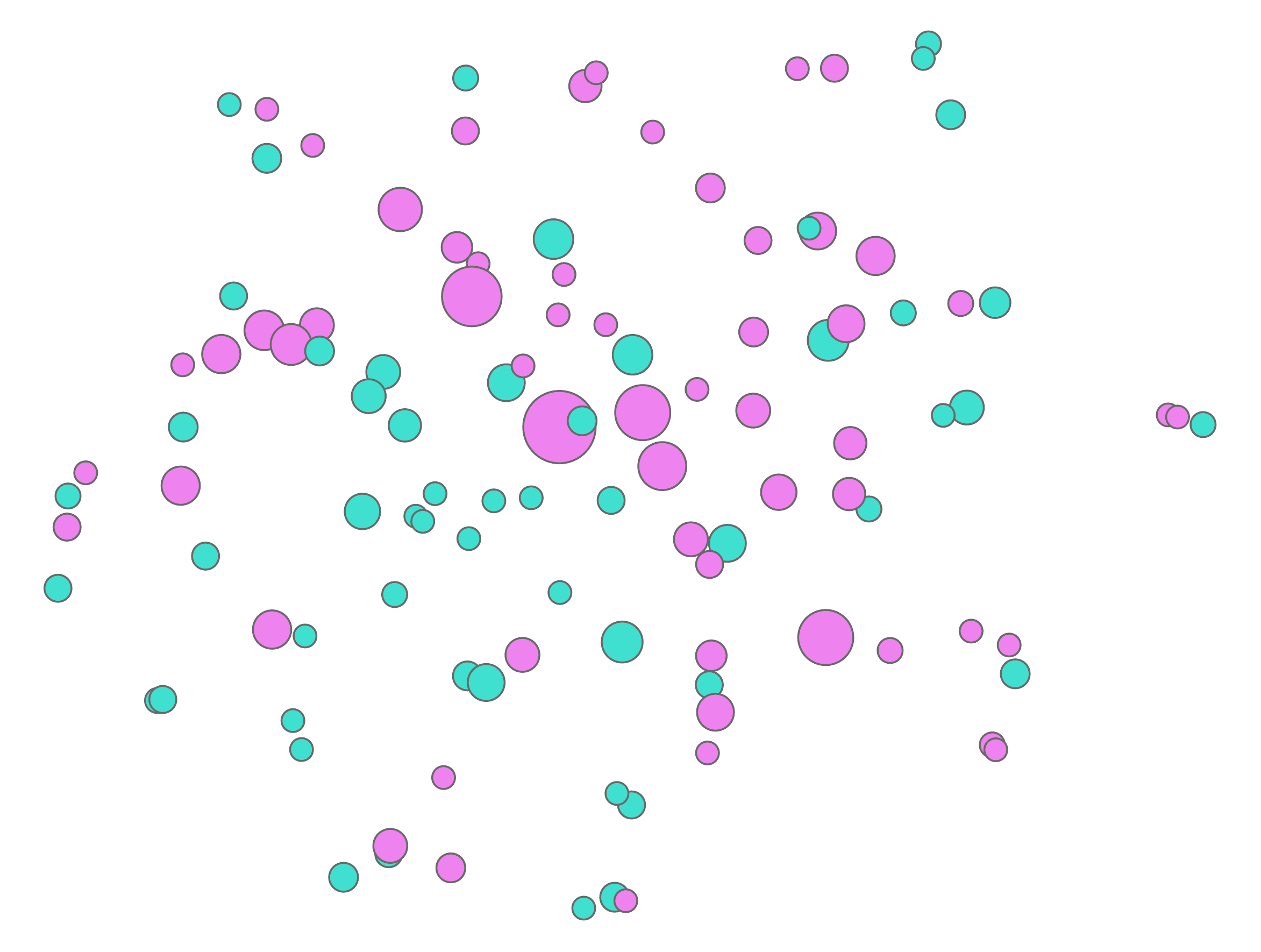}\label{fig:adjnoun-r0_5}}
\subfloat[$r=0.3$]{\includegraphics[width=0.23\textwidth, keepaspectratio]{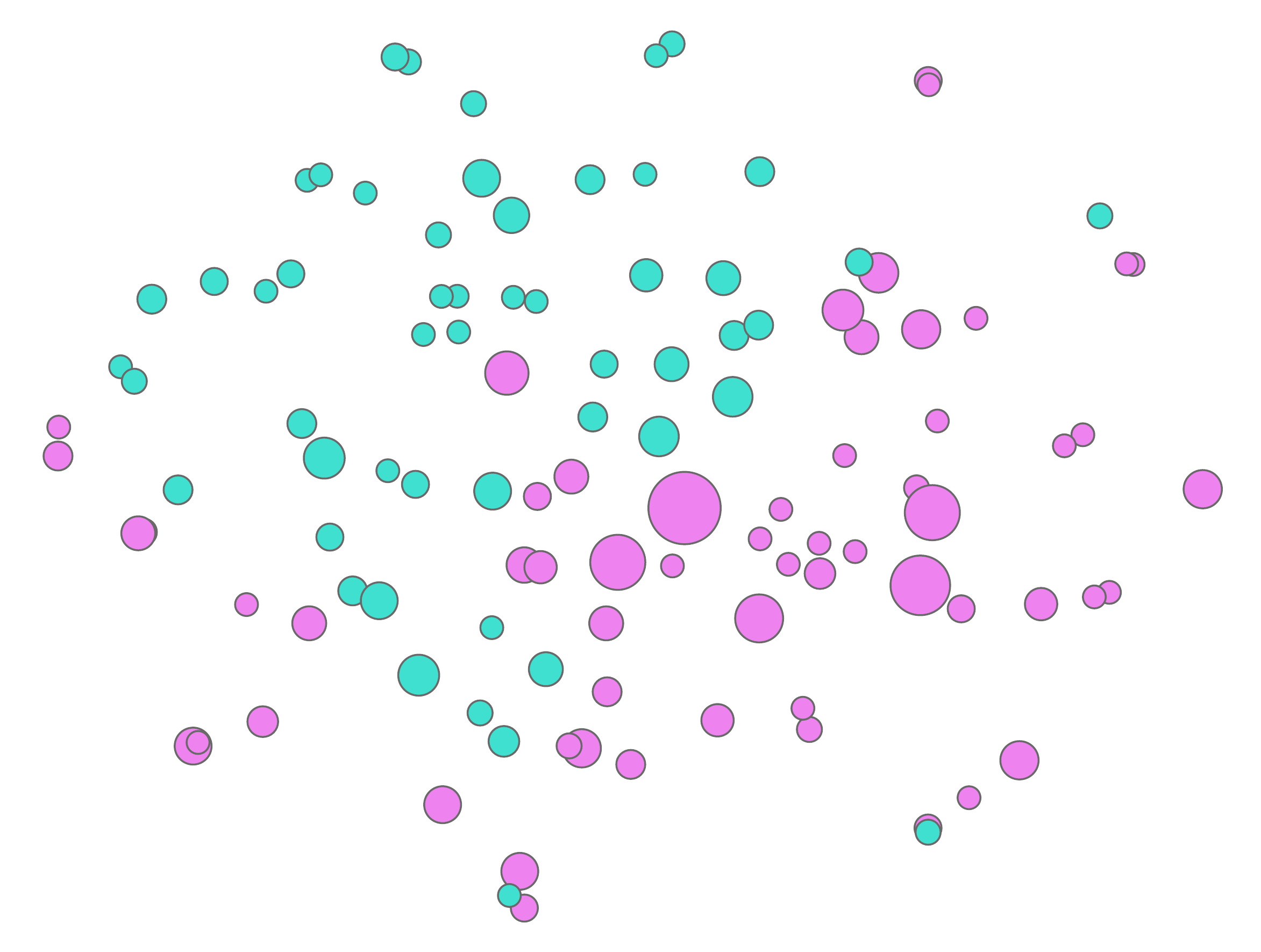}\label{fig:adjnoun-r0_3}}
\subfloat[$r=0.1$]{\includegraphics[width=0.23\textwidth, keepaspectratio]{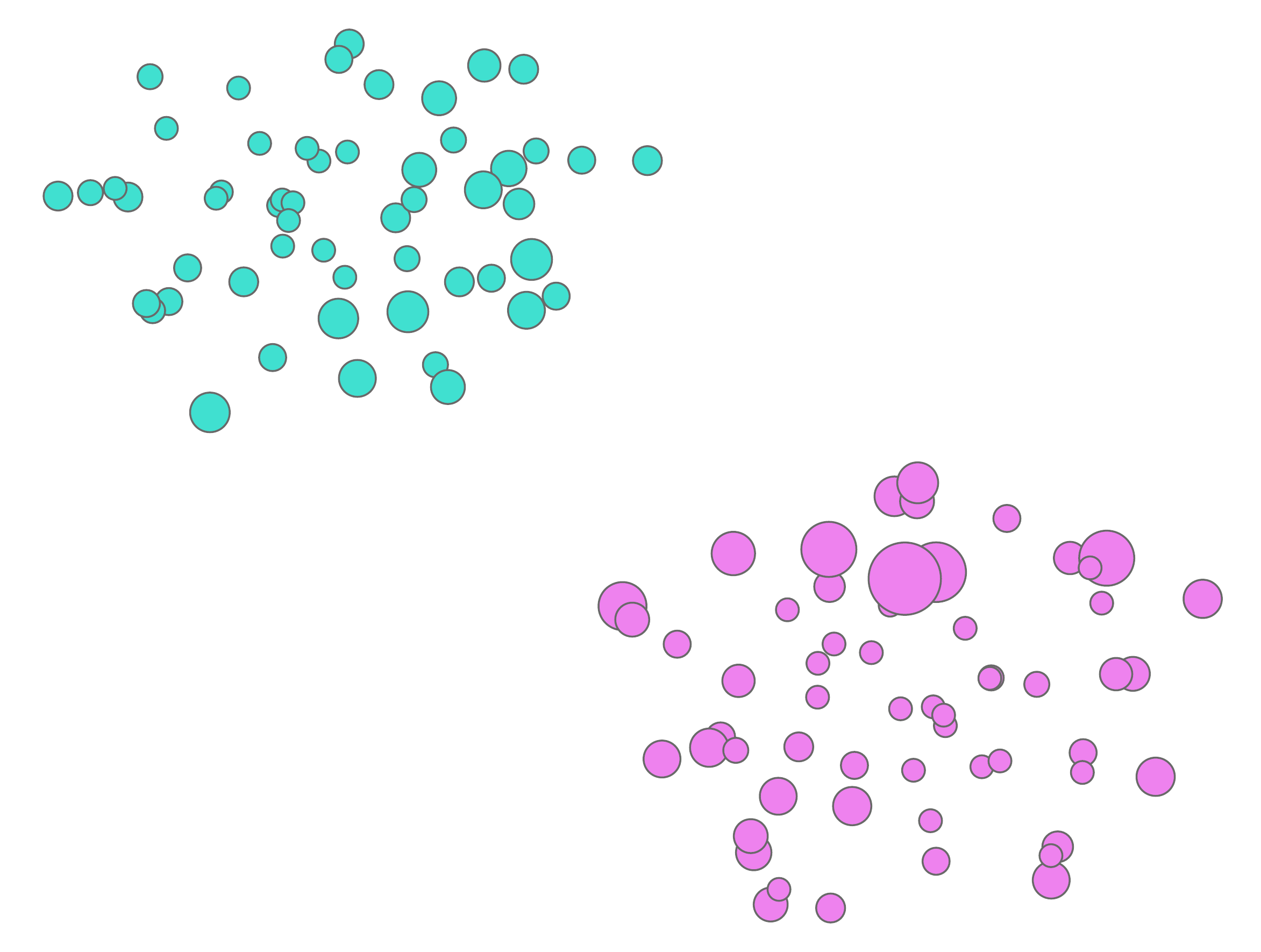}\label{fig:adjnoun-r0_1}}
\caption[]{Illustration of different $r$ in FANE with Dataset Adjnoun \cite{RN20144}. Node colors represent different property attributes of English words. As we can see, with smaller $r$, the embedding results in more attribute homophily.}
\label{fig:adjnoun-r}
\end{figure*}

Let the constructed network be denoted as $G' = (V', E', W', \Lambda')$, where $V'$ is the set of raw nodes and virtual attribute node, $E'$ is the set of raw edges and virtual edges between nodes and corresponding attribute nodes, $\Lambda'$ is same with $\Lambda$ besides node properties of attribute nodes. Each attribute node $v_{\alpha_i} \in V'$ is associated with an attribute vector where only $\alpha_i(v_{\alpha_i})$ is nonzero. $W'$ includes $W$ and weights for attribute edges, which can be defined as needed. The objective function evolves as:
\begin{equation}
  \prod_{u \in V'} P'(N_s'(u)|f'(u))
\end{equation}
So the aim is to compute the new mapping function $f'$. It is interesting that we get an side effect that network properties can be embedded too. So we could also learn features specifically on network property. It is non-trial as shown in experiments in Section \ref{sec:experiments}.

\subsection{Context generation}
After network construction, we can integrate property information into various state-of-the-art structure-preserving network embedding methods. However, We confront the second key challenge: how to provide the flexibility (T3) in various situations. That is to say, how FANE support user get structure- or property-preserving feature learning or both as needed. Inspired by node2vec \cite{RN20017}, we designed a new random walking strategy which allows continuous transition between attribute-preserving and structure-preserving network embedding.

Given a source node $v \in V'$ and a fixed random walking length $l$. The $i$th node in the walk $c_i$, which starts with $c_0 = u$, is generated by the following distribution:
\begin{equation}
  P(c_i=x|c_{i-1}=v)=
  \begin{cases}
      \frac{\pi_{vx}}{Z},& \text{if } (v, x)\in E'\\
      0,              & \text{otherwise}
  \end{cases}
\end{equation}
where $\pi_{vx}$ is the unnormalized transition probability between nodes $v$ and $x$, and $Z$ is the normalizing constant. Notice that $E'$ includes constructed attribute edges.

As illustrated in Figure \ref{fig:Attri2Vec}, a random walking that just traversed to node $v$ through edge $(u, v)$. Notice that node $v$ and $x_2$ have same attribute $a_2$, so there will be two more \textit{virtual} edges connecting the \textit{virtual} attribute node $a_2$ with node $v$ and with node $x_2$ in network $G'$ separately. One key here is how to define the probability for attribute node related walking. In the next section, we will exhaustively discuss three possible random walking strategies and analyze the effectiveness following our goal (T3).

\subsection{Random walking strategies}
After enumeration case by case, we find that there are three different random walking strategies in the constructed attributed network $G'$:
\begin{enumerate}
\item Source Focused (FANE-sf). Let $V_\Lambda \subset V'$ denote the set of attribute nodes in $G'$ and let $\alpha(v, x)$ denotes the probability of the next step from the source node $v$ to the target node $x$. And $\pi_{vx}=w'(v, x)*\alpha(v, x)$. The probability, $\alpha(v, x)$, is defined as $1/r$ only if the source node $v \in V_\Lambda$. In practice, there are four conditions as shown in Figure \ref{fig:FANE-sf}. Formally, the probability of the next step is designed as:
\begin{equation}
  \alpha(v, x)=
  \begin{cases}
      \frac{1}{r},& \text{if } v \in V_\Lambda\\
      \beta(v, x),& \text{otherwise}
  \end{cases}
\end{equation}
where $d_{vx}$ denotes the shortest path distance between nodes $v$ and $x$, and
\begin{equation} \label{eq:disvx}
  \beta(v, x)=
  \begin{cases}
      \frac{1}{p},& \text{if } d_{ux}=0\\
      1,          & \text{if } d_{ux}=1\\
      \frac{1}{q},      & \text{if } d_{ux}=2\\
  \end{cases}
\end{equation}
where $u \in V'$ which is the previous node of $v \in V'$.
\item Target Focused (FANE-tf). As illustrated in Figure \ref{fig:FANE-tf}, the probability is defined as $1/r$ only if the target node $x \in V_\Lambda$. The probability of the next step is designed as:
\begin{equation}
  \alpha(v, x)=
  \begin{cases}
      \frac{1}{r},& \text{if } x \in V_\Lambda\\
      \beta(v, x),& \text{otherwise}
  \end{cases}
\end{equation}
where $\beta(v, x)$ is the same as equation \ref{eq:disvx}.
\item Source \& Target Focused (FANE-stf). As illustrated in Figure \ref{fig:FANE-stf}, the probability is defined as $1/r$ if the source or target node $v \text{ or } x \in V_\Lambda$. The probability of the next step is designed as:
\begin{equation}
  \alpha(v, x)=
  \begin{cases}
      \frac{1}{r},& \text{if } v \text{ or } x \in V_\Lambda\\
      \beta(v, x),& \text{otherwise}
  \end{cases}
\end{equation}
Again, $\beta(v, x)$ is the same as equation \ref{eq:disvx}.
\end{enumerate}
 
Under our proposed framework, the random walking should be able to biased toward the attribute nodes in order to be attribute-preserving. However, the probability of random walking from the source node to the attribute node, by definition, depends on the values of $p$, $q$ and $d_{ux}$. Consequently, it is hard to define the probability of random walking such that the walking is always biased toward the attribute node. So strategy FANE-sf could be structure-preserving by adjusting the values of $p$ and $q$, it is unlikely to be attribute-preserving. This result contradicts with our objectives. Experiment also confirms our analysis as shown in Figure \ref{fig:attri2vec-f}, in which strategy FANE-sf yields little separation between two subsets of nodes with different attributes.

In strategies FANE-sf and FANE-stf, the probability of property-preserving is defined by $r$. By adjusting its values, it is possible to make the random walking be structure-preserving or attribute-preserving. We will thoroughly discuss the effect of $r$ in Section \ref{sec: attripara}. This result satisfies our objectives. In Figure \ref{fig:attri2vec-f}, we can see that by setting $r$ smaller than $p$ and $q$, the biased random walking is attribute-preserving, both strategies yield distinct separation between two different attributes. We choose strategy FANE-tf biased random walking strategy in the following experiments.

\subsection{Attribute Parameter, $r$} \label{sec: attripara}
While $p$ and $q$ control the likelihood of walking to local and structure nodes separately \cite{RN20017}, property parameter $r$ decides the bias between structure- and property-preserving. Let us consider two cases:
\begin{itemize}
\item Case 1: $r > max(p, q, 1)$
By setting $r$ to be large, it is unlikely for the source node walking to the attribute nodes during the random walking process. Thus, the random walking is more biased toward preserving the local and the structural information of the network based on the values of $p$ and $q$.
\item Case 2: $r < min(p, q, 1)$
By setting $r$ to be small, we increase the probability of walking from the source nodes to the attribute nodes. Thus, nodes sharing similar attribute information are more likely to be linked together by the attribute nodes. Consequently, the result of network embedding will be more property-preserving. Figure \ref{fig:adjnoun-r} validates the embedding effectiveness of FANE with different $r$ on real datasets. As we can see, the embedding result will be more property homophily by decreasing $r$.
\end{itemize}
We can see that the hyper-parameter $r$ functions as a slider between structure- and attribute-preserving. In Section 4, we will conduct an experiment to elaborate how our method can integrate both structure-preserving and property-preserving feature learning (T2) and make continuous transition between both (T3).

\subsection{Algorithm}\label{Algorithm}
The pseudocode of FANE is shown in Algorithm \ref{algo:FANE}. We first construct an property-enhanced network $G'$. By importing attribute node in-out hyperparameter $r$, we could control the weights of structure-preserving and property-preserving random walking. The source code and datasets of this paper can be obtained from \url{https://github.com/GraphWorld/FANE}.

\begin{algorithm}
    \SetKwInOut{Input}{Input}
    \SetKwInOut{Output}{Output}
    \SetKwProg{Fn}{faseWalk}{:}{}
    \Input{Network $G = (V, E, \Lambda)$, Dimensions $d$, Walks per
node $w$, Walk length $l$, Context size $k$, Return $p$, Raw node in-out $q$, Attribute node in-out $r$}
    \Output{Embedding matrix $f \in \mathbb{R}^{|\Lambda' \times d|}$}
    $G'$ = $G$\;
    \For{$x \in \Lambda$}{
      Append node $a_x$ to $G'$\;
      \For{$v \in V$}{
          \If{$v$ has attribute $x$}{
              Append edge $(v, a_x)$ to $G'$\;
          }
      }
    }
    $\pi$ = PreprocessModifiedWeights($G', p, q, r$)\;
    $G'$ = $(V', E', \pi)$\;
    Initialize \textit{walks} to Empty\;
    \For{iter = 1 to w}{
        \For{$u \in V'$}{
            $walk$ = faseWalk($G', u, l$)\;
            Append $walk$ to $walks$\;
        }
    }
    f = StochasticGradientDescent($k, d, walks$)\;
    return $f$\;
    \caption{FANE}\label{algo:FANE}
\end{algorithm}
\section{Experiments} \label{sec:experiments}
The proposed method is flexible enough for embedding of network datasets from different domains as discussed below.

\subsection{Datasets}
We evaluate our method on several datasets from different domains, as listed in Table \ref{tab:datasets}.

\begin{itemize}
\item Adjnoun \cite{RN20144}: Nodes represent the most commonly occurring adjectives and nouns in novel David Copperfield. Edges connect any pair of words that occur in adjacent position. The property attributes, adjectives and nouns, are used as node classification information.
\item WebKB \cite{RN20148}: Nodes and edges represent web-pages and citation network separately. Attributes are described as 0/1-valued word vectors indicating the absence/presence of the corresponding word from the dictionary.
\item Cora \cite{RN20148}: Nodes represent scientific publications and edges reflect the citation relationships. Similar with dataset WebKB, attributes are described as 0/1-valued word vectors indicating the absence/presence of the corresponding word from the dictionary.
\item CiteSeer \cite{RN20147}: Nodes represent scientific publications and edges reveal the citation networks. Similarly, attributes are described as 0/1-valued word vectors indicating the absence/presence of the corresponding word from the dictionary.
\item ego-Facebook \cite{RN20118}: Nodes represent Facebook survey participants. The friend list is shown as edge. Attributes are anonymous personal information of those participants.
\item ego-Gplus \cite{RN20118}: Nodes represent Google users who decide to share their circles. The friend list is shown as edge too. Attributes are their personal information.
\item ego-Twitter \cite{RN20118}: Nodes are participants from Twitter. Edge represent following list. Attributes are hashtags or user themselves in twitter.
\end{itemize}

\begin{table}
\centering
  \caption{Statistics of the datasets}
  \label{tab:datasets}
  \begin{tabular}{l|r|r|r|r}
    \toprule
    \multicolumn{1}{c|}{Name} & \multicolumn{1}{c|}{$V$} & \multicolumn{1}{c|}{$E$} & \multicolumn{1}{c|}{$\Lambda$} & \multicolumn{1}{c}{$G$} \\ \hline
    adjnoun & 112 & 425 & 2 & - \\ \hline
    WebKB & 877 & 1,608 & 1,703 & 5 \\ \hline
    Cora & 2,708 & 5,278 & 1,433 & 7 \\ \hline
    CiteSeer & 3,312 & 4,732 & 3,703 & 6 \\ \hline
    ego-Facebook & 4,039 & 88,234 & 1,283 & 191 \\ \hline
    ego-Gplus & 7,856 & 321,268 & 2,024 & 91 \\ \hline
    ego-Twitter & 2,511 & 37,154 & 9,073 & 132 \\
    \bottomrule
  \end{tabular}
\end{table}

\begin{figure*}
\centering
\subfloat[node2vec \cite{RN20017}]{\includegraphics[width=0.3\textwidth, keepaspectratio]{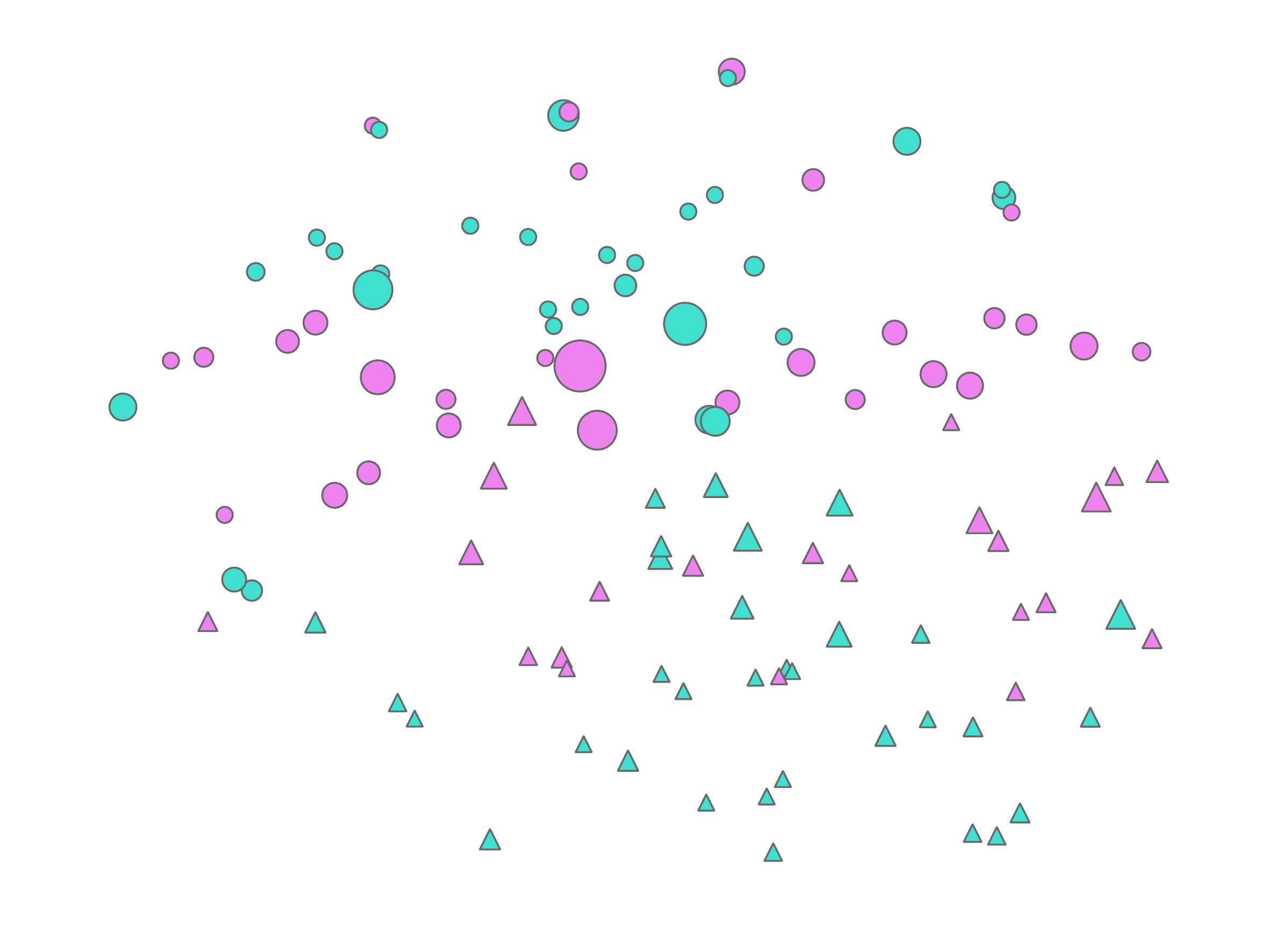}\label{fig:c-node2vec}}
\subfloat[struc2vec \cite{RN20124}]{\includegraphics[width=0.3\textwidth, keepaspectratio]{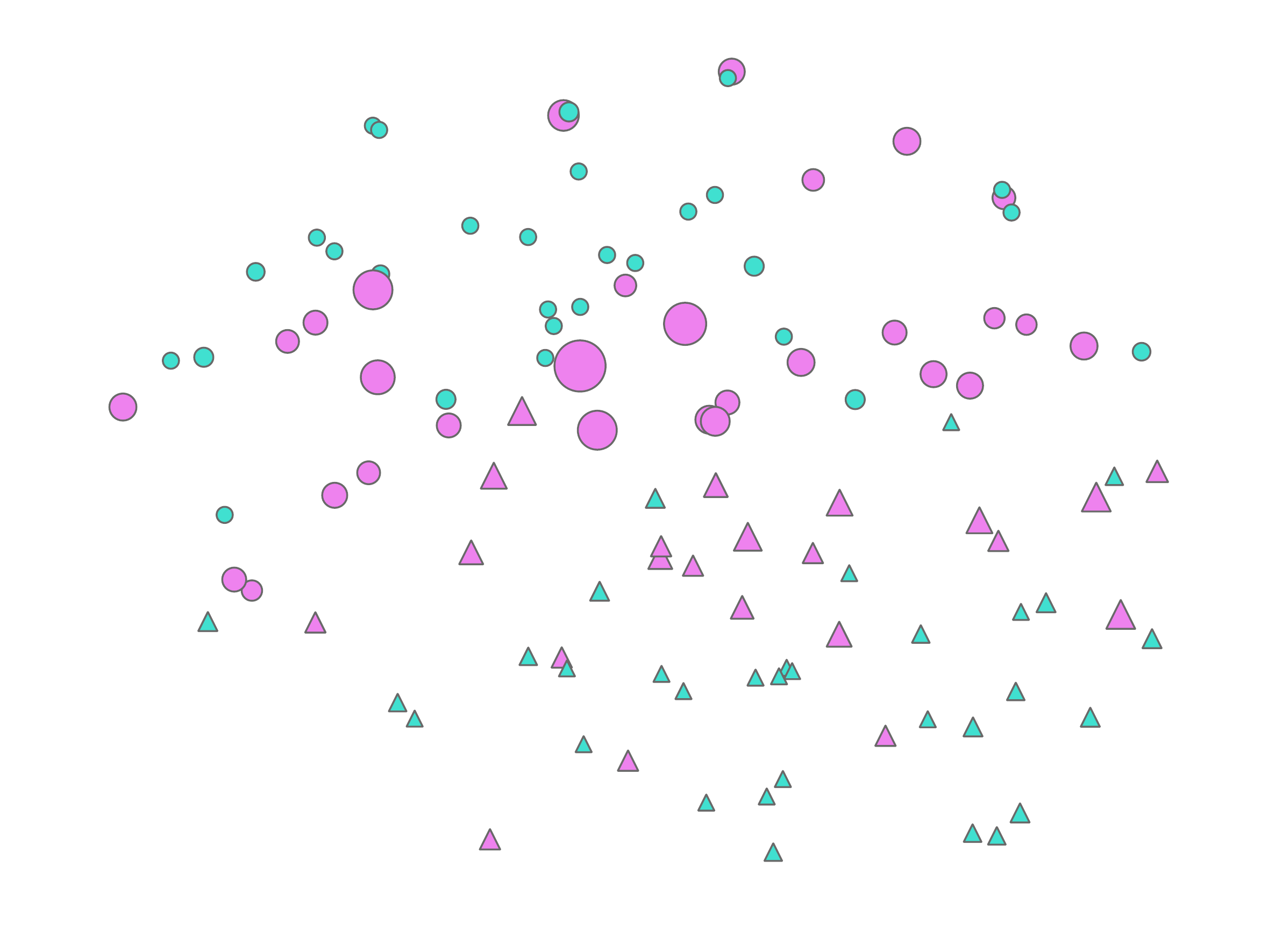}\label{fig:c-struc2vec}}
\subfloat[FANE with $r=10.0$]{\includegraphics[width=0.3\textwidth, keepaspectratio]{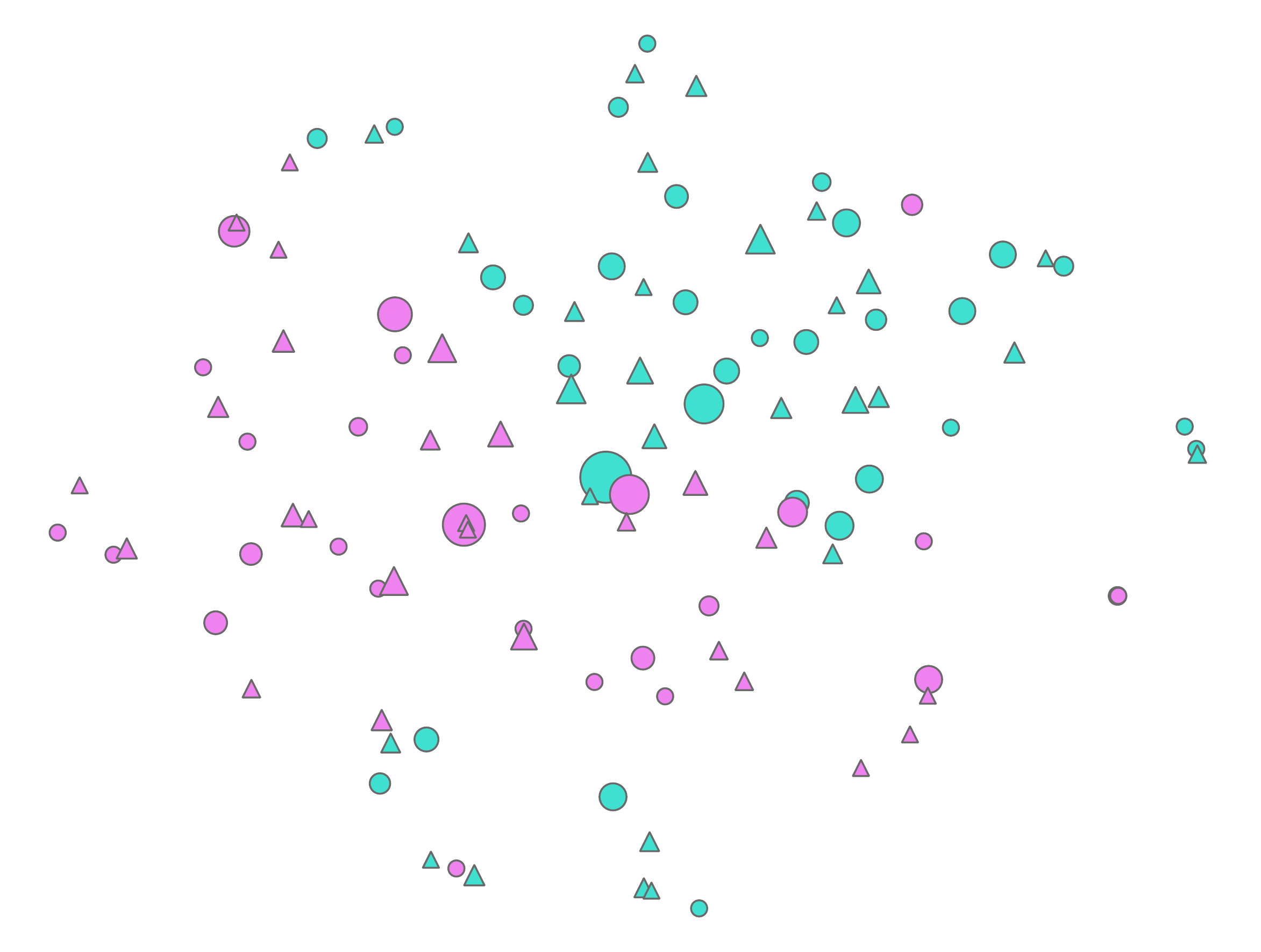}\label{fig:c-attri2vec10}} \newline
\subfloat[TADW \cite{RN20134}]{\includegraphics[width=0.3\textwidth, keepaspectratio]{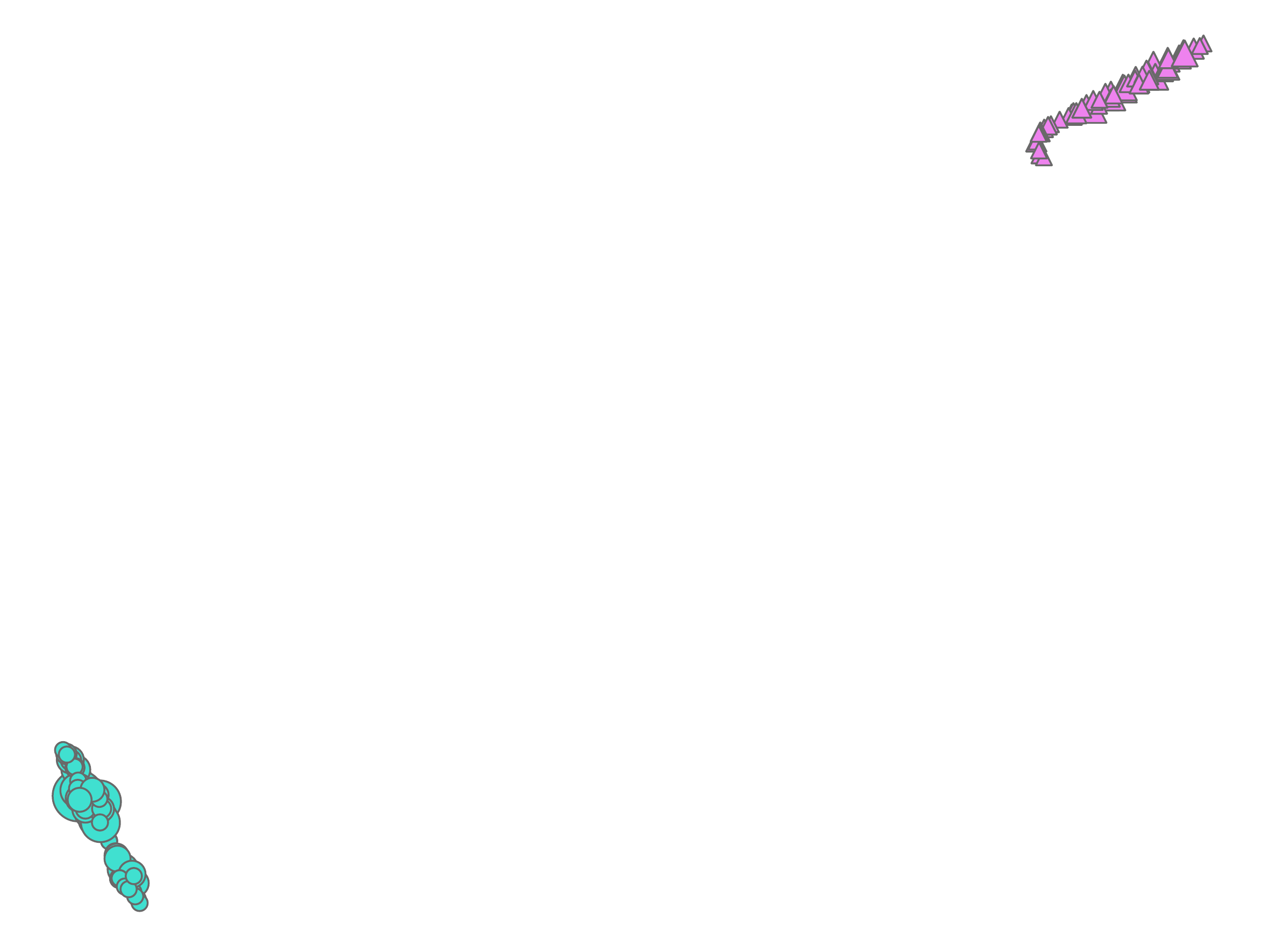}\label{fig:c-TADW}}
\subfloat[HSCA \cite{RN20136}]{\includegraphics[width=0.3\textwidth, keepaspectratio]{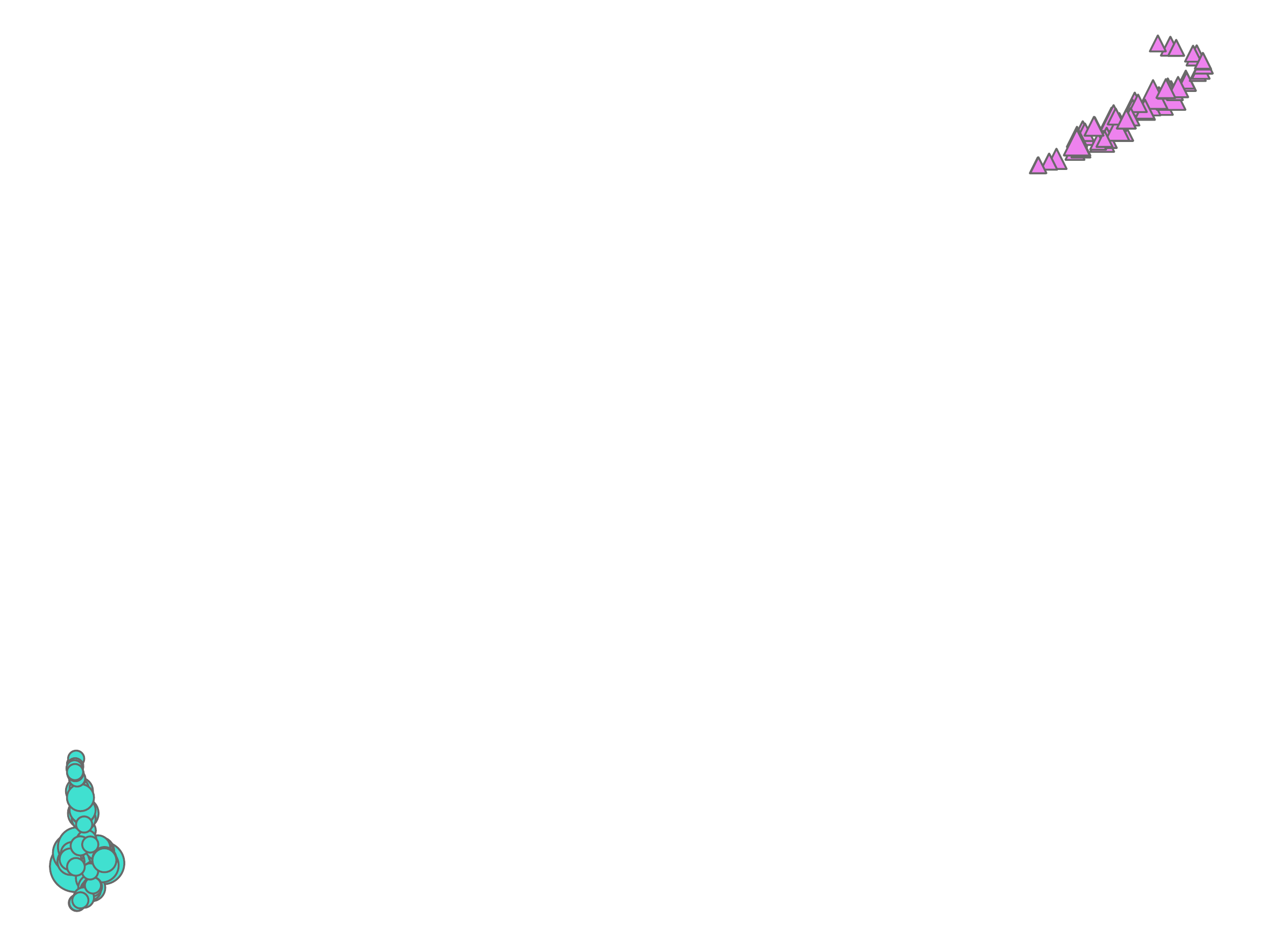}\label{fig:c-HSCA}}
\subfloat[FANE with $r=0.01$]{\scalebox{-1}[1]{\includegraphics[width=0.3\textwidth, keepaspectratio]{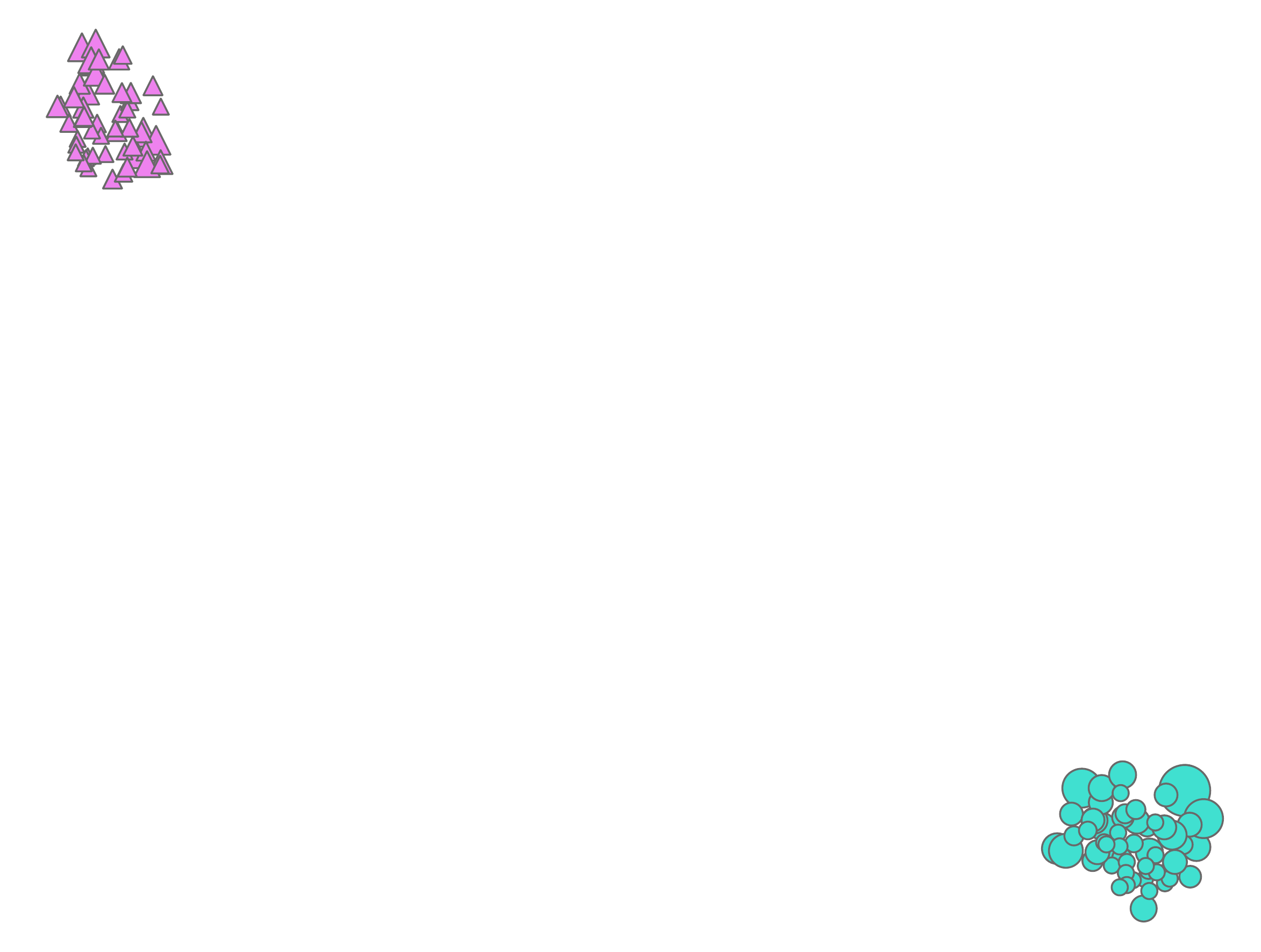}\label{fig:c-attri2vec001}}} \newline
\subfloat[FANE with $r=0.3$]{\includegraphics[width=0.3\textwidth, keepaspectratio]{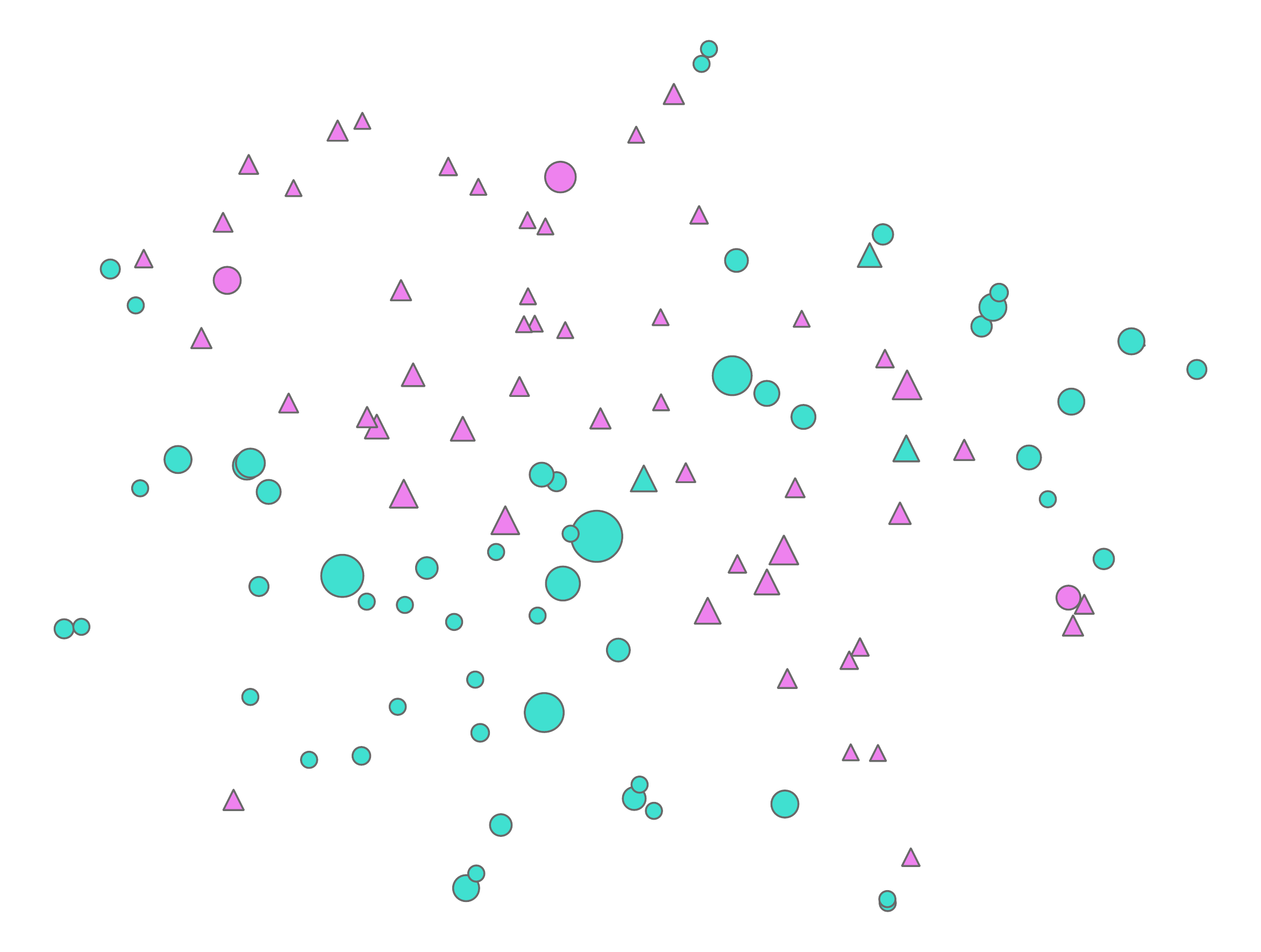}\label{fig:c-attri2vec03}}
\subfloat[FANE with $r=0.2$]{\includegraphics[width=0.3\textwidth, keepaspectratio]{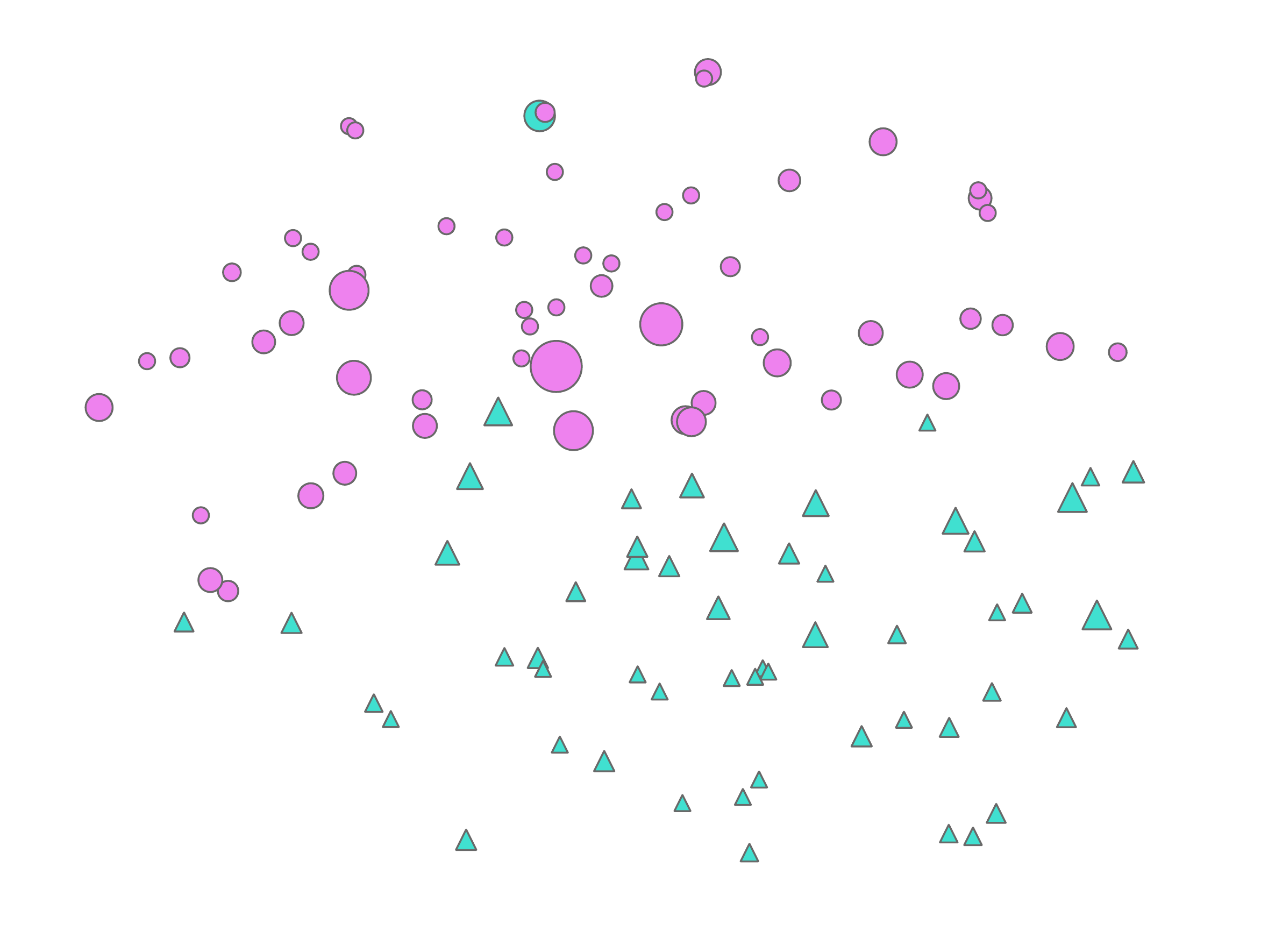}\label{fig:c-attri2vec02}}
\subfloat[FANE with $r=0.1$]{\includegraphics[width=0.3\textwidth, keepaspectratio]{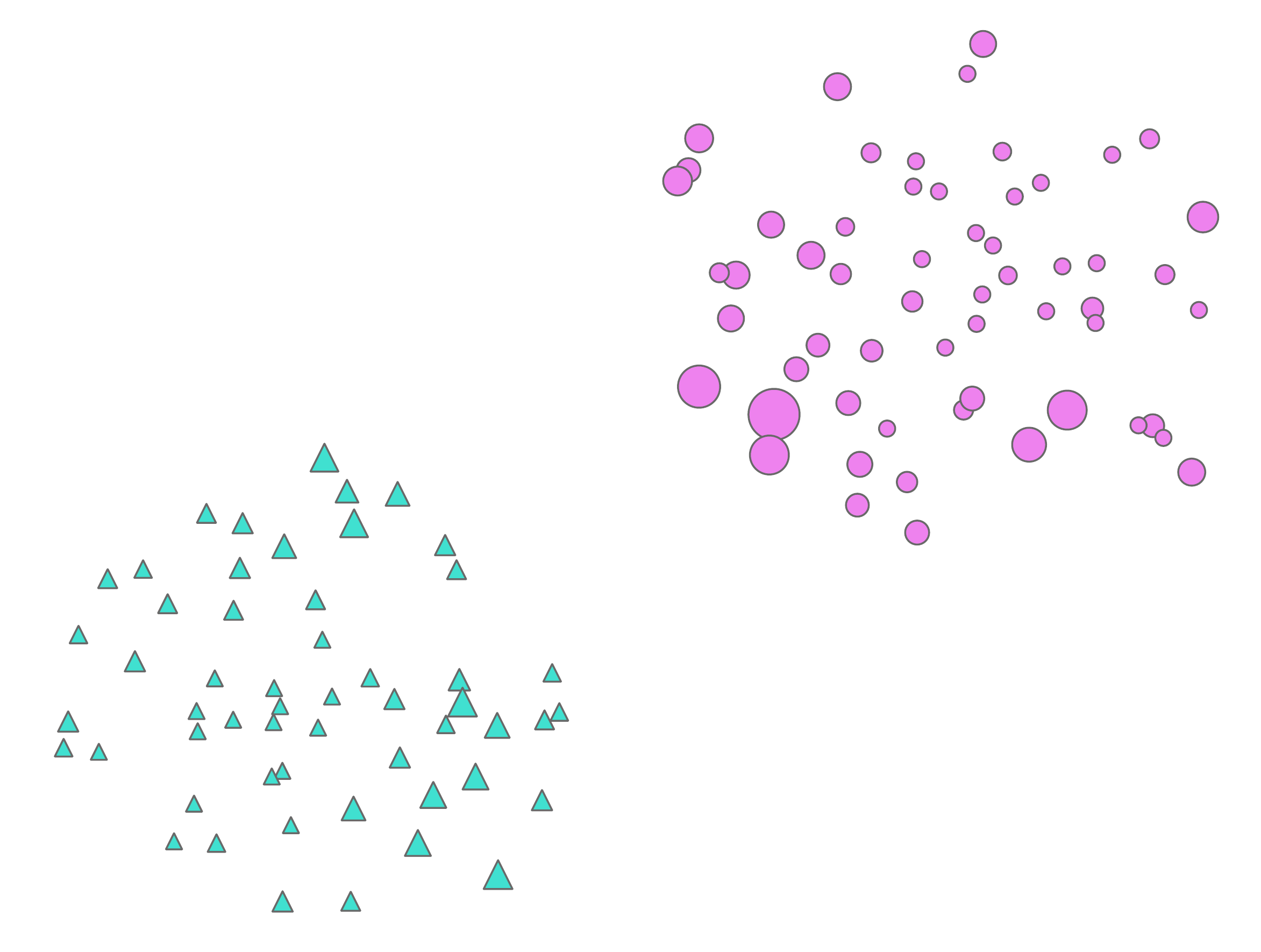}\label{fig:c-attri2vec01}}
\caption[]{Illustrations of FANE, which is flexible enough to get results as STAR structure-preserving and attribute-preserving network embedding methods with Dataset Adjnoun \cite{RN20144}. Node shapes and colors represent different property attributes and clustering results of English words separately. As we can see, structure-preserving methods, node2vec (a) and struc2vec (b), represent more structure homophily, attribute-preserving, TADW (d) and HSCA (e), represent extreme attribute homophily. Our method provides flexibility in integrating structure homophily and attribute homophily, which could learn structure-preserving features (c) as node2vec and struc2vec and attribute-preserving features (f) as TADW and HSCA. Moreover, FANE provides the ability of smoothly transiting between structure homophily and attribute homophily (g-i).}
\label{fig:comparation}
\end{figure*}

\subsection{Baselines}
We compare FANE with several state-of-the-art network embedding methods. The implements of these methods are from the original authors.
\begin{itemize}
\item node2vec \cite{RN20017}: This approach provide a way to integrate Breadth-first Sampling and Depth-first Sampling in random walking, which introduces the Skip-Gram algorithm to learn the node representation vectors.
\item struc2vec \cite{RN20124}: This approach first encodes the vertex structural role similarity into a multilayer network. The weights of edges at each layer are determined by the structural role difference at the corresponding scale.
\item TADW \cite{RN20134}: This approach extends DeepWalk \cite{RN20110} by encoding node text features into the matrix factorization.
\item HSCA \cite{RN20136}: This approach simultaneously integrates homophily, structural context, and vertex content to learn effective network representations.
\end{itemize}

\begin{figure*}
\centering
\includegraphics[width=0.9\textwidth]{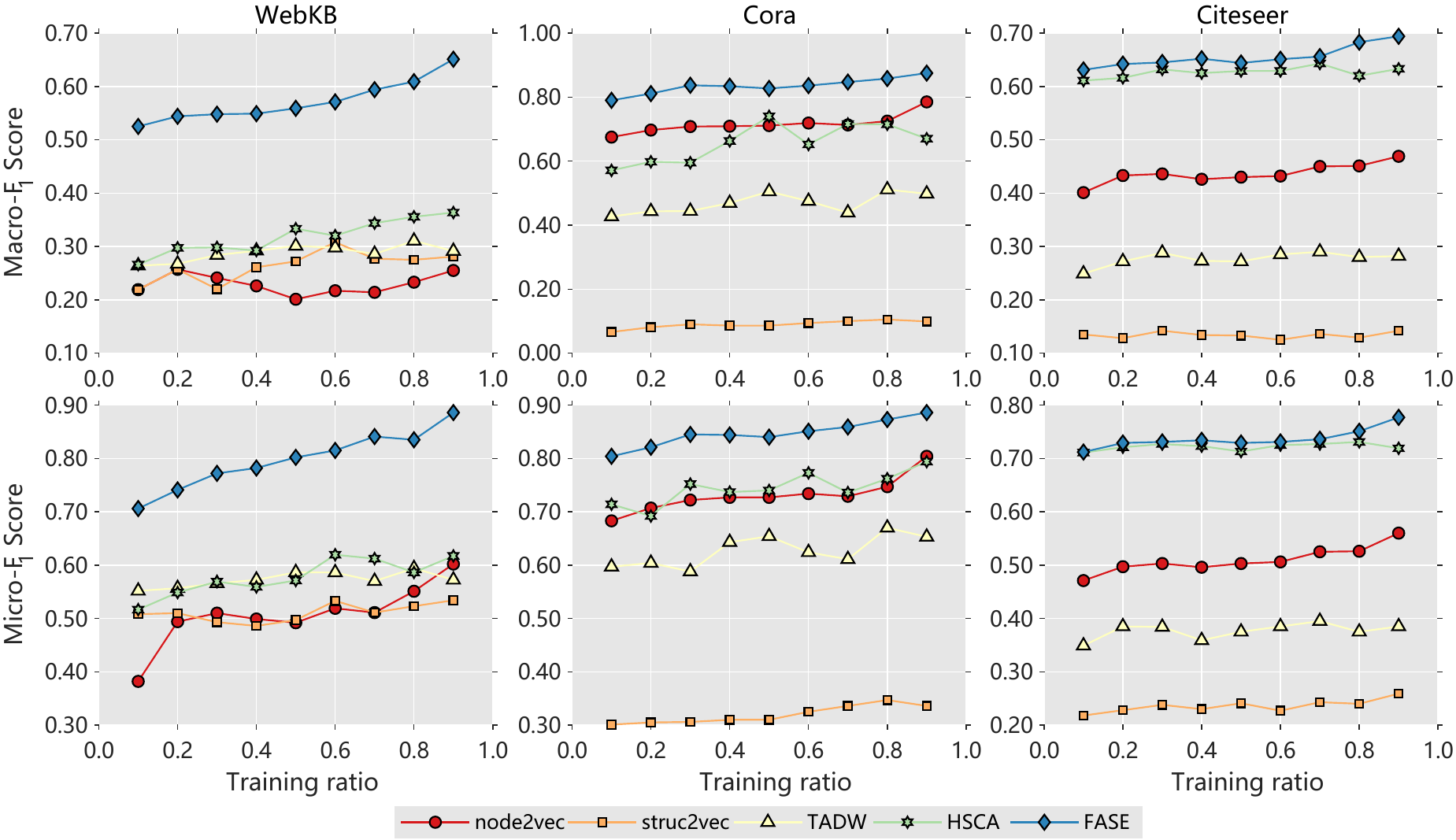}
\caption{Evaluation results (higher is better) on WebKB, Cora and Citeseer datasets.}
\label{fig:classification}
\end{figure*}

\subsection{Case Study: Adjnoun}
Different with existing methods, FANE could realize flexible structure homophily and attribute homophily. Firstly, we empirically test our embedding results in comparison to structure homophily methods, node2vec($p$ = $q$ = 1) and struc2vec, and attribute homophily methods, TADW($d$ = 8, textRank = 20, lambda = 0.2, train\_ratio = 0.5, $C$ = 5) and HSCA($d$ = 8, textRank = 20, lambda = 0.2, train\_ratio = 0.5, mu = 0.1, $C$ = 5) as shown in Figure \ref{fig:comparation}. Notice that we set the embedding dimension, $d$, and textRank to be low for TADW and HSCA given the small node size of dataset adjnoun.

To begin with, our method can be either mainly structural homophily or attribute homophily by tuning the values of $r$. In Figure \ref{fig:c-node2vec} to \ref{fig:c-attri2vec10}, we can see that by setting $r$ to be large, FANE can yield similar result as those of node2vec and struc2vec in terms of preserving structural relationship. In addition, in Figure \ref{fig:c-TADW} to \ref{fig:c-attri2vec001}, when $r$ is set to be small, the embedding result of FANE can also be property-preserving like those of TADW and HSCA. Moreover, FANE can extract aptitudinal features reflecting both structure and attribute homophily by adjusting the hyper-parameter $r$, as shown in Figure \ref{fig:c-attri2vec01} to \ref{fig:c-attri2vec03} (T2). When $r = 0.1$, the embedding result integrates relatively more attribute information while for $r = 0.3$, relatively more structural information is preserved. For further explain the effects of FANE in integrating structure and attribute homophily, we take $r=0.2$ as an example, as shown in Figure \ref{fig:c-attri2vec02}, the result classes (shown with different colors) are mostly consistent with property attributes (shown with different shapes). Notice that in this situation, there is still one node (Green circle) which represent adjective word \enquote{first} is classified with nouns. The reason is that the word are connected closely with nouns in structure. It proves that FANE preserves structure homophily even in this extreme setting (T3).

The Word network is an effective example demonstrating the functionality of our proposed method. In the following section, we will conduct additional experiments to evaluate the effectiveness of our embedding method on network classification.

\begin{figure*}
\centering
\subfloat[node2vec]{\includegraphics[width=0.2\textwidth]{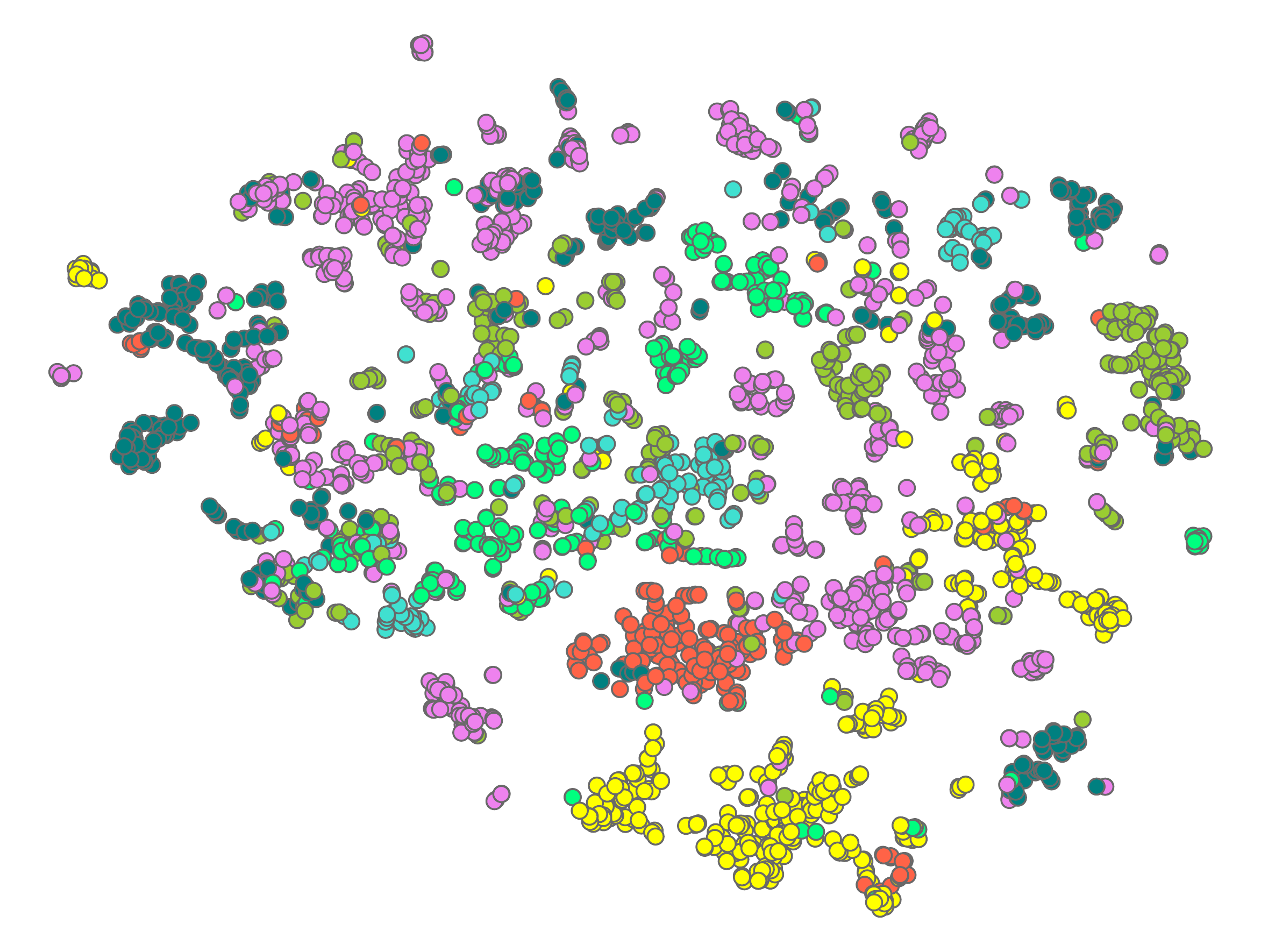}\label{fig:cora-n}}
\subfloat[struc2vec]{\includegraphics[width=0.2\textwidth]{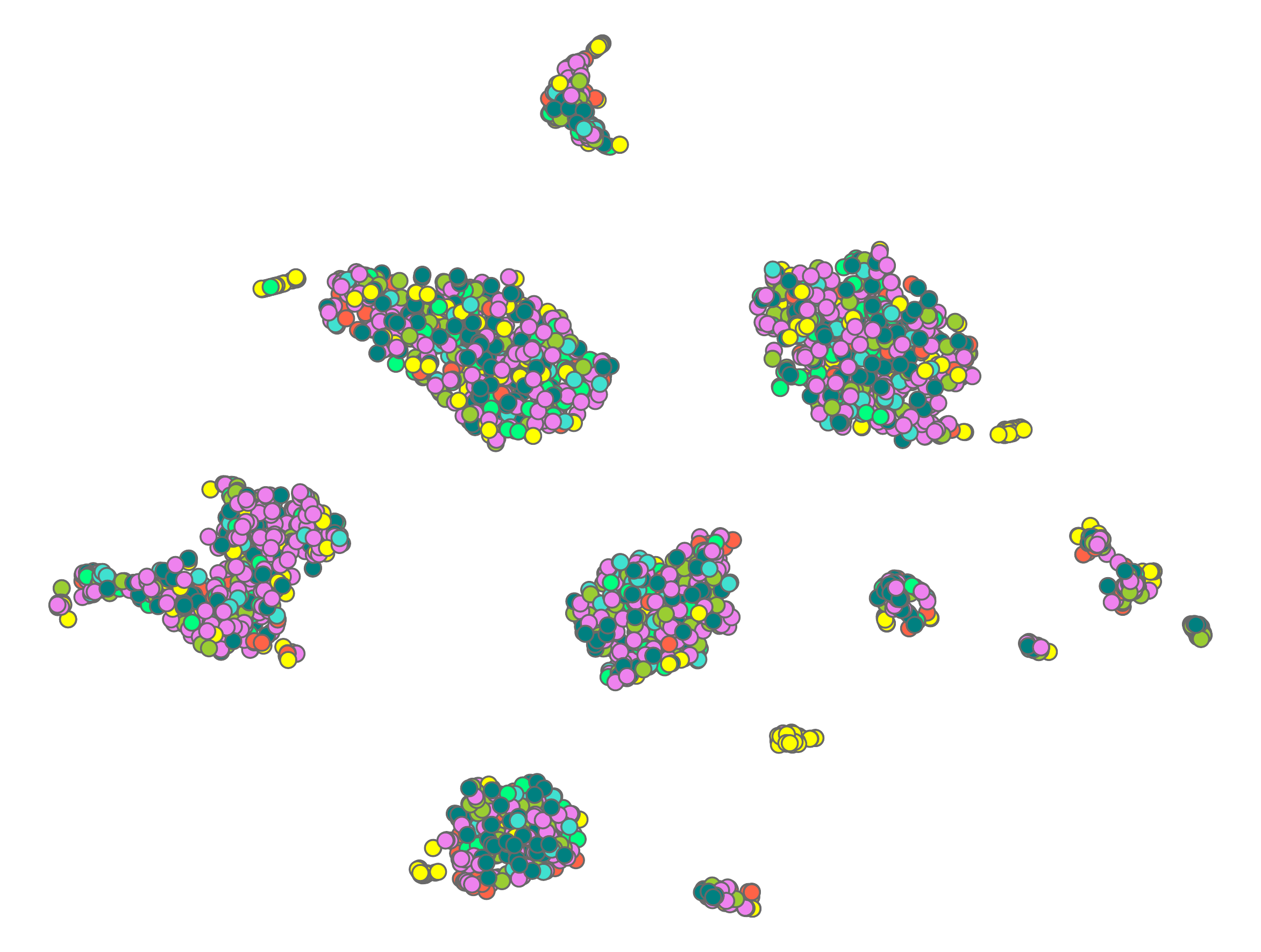}\label{fig:cora-s}}
\subfloat[TADW]{\includegraphics[width=0.2\textwidth]{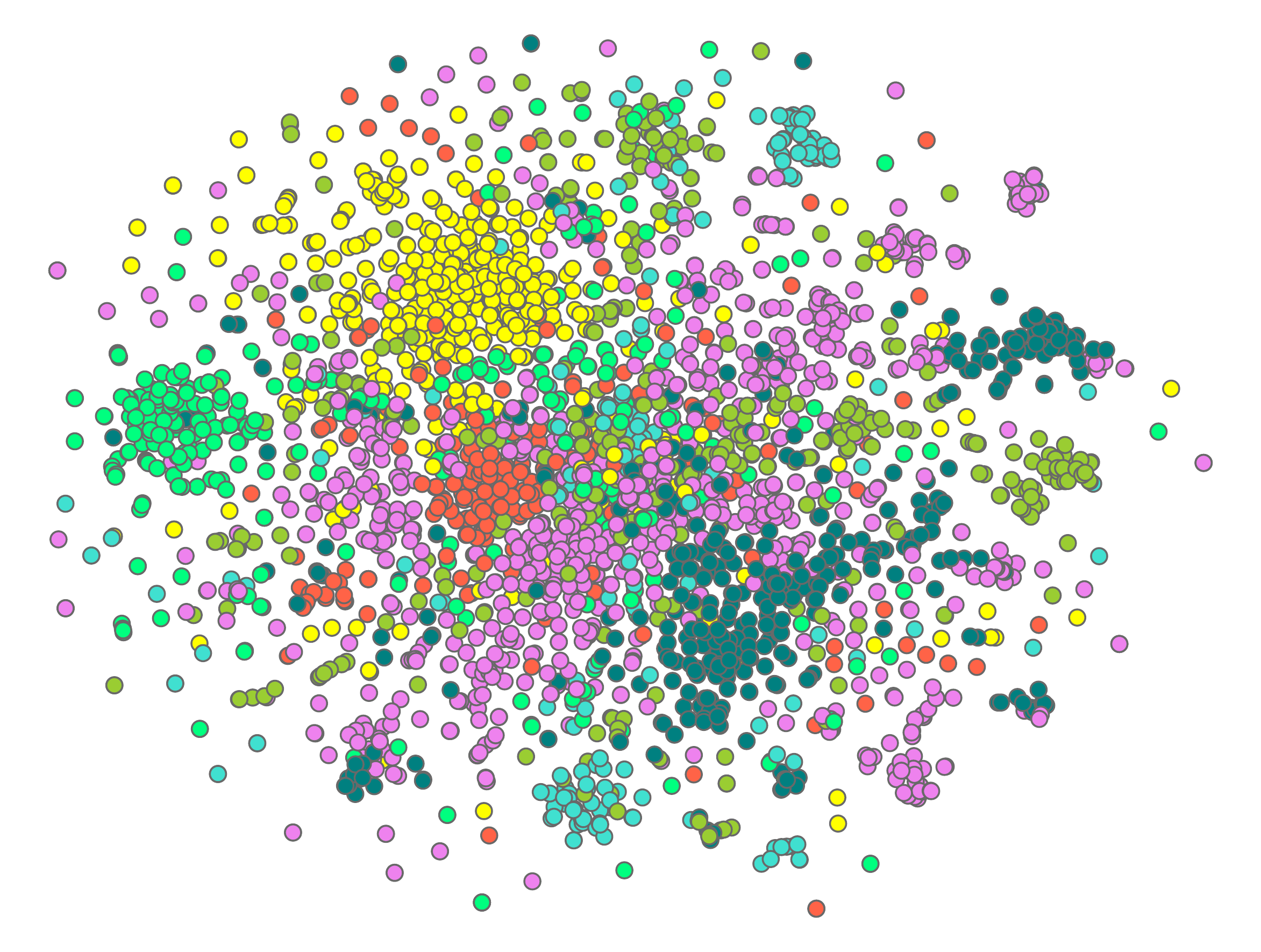}\label{cora-t}}
\subfloat[HSCA]{\includegraphics[width=0.2\textwidth]{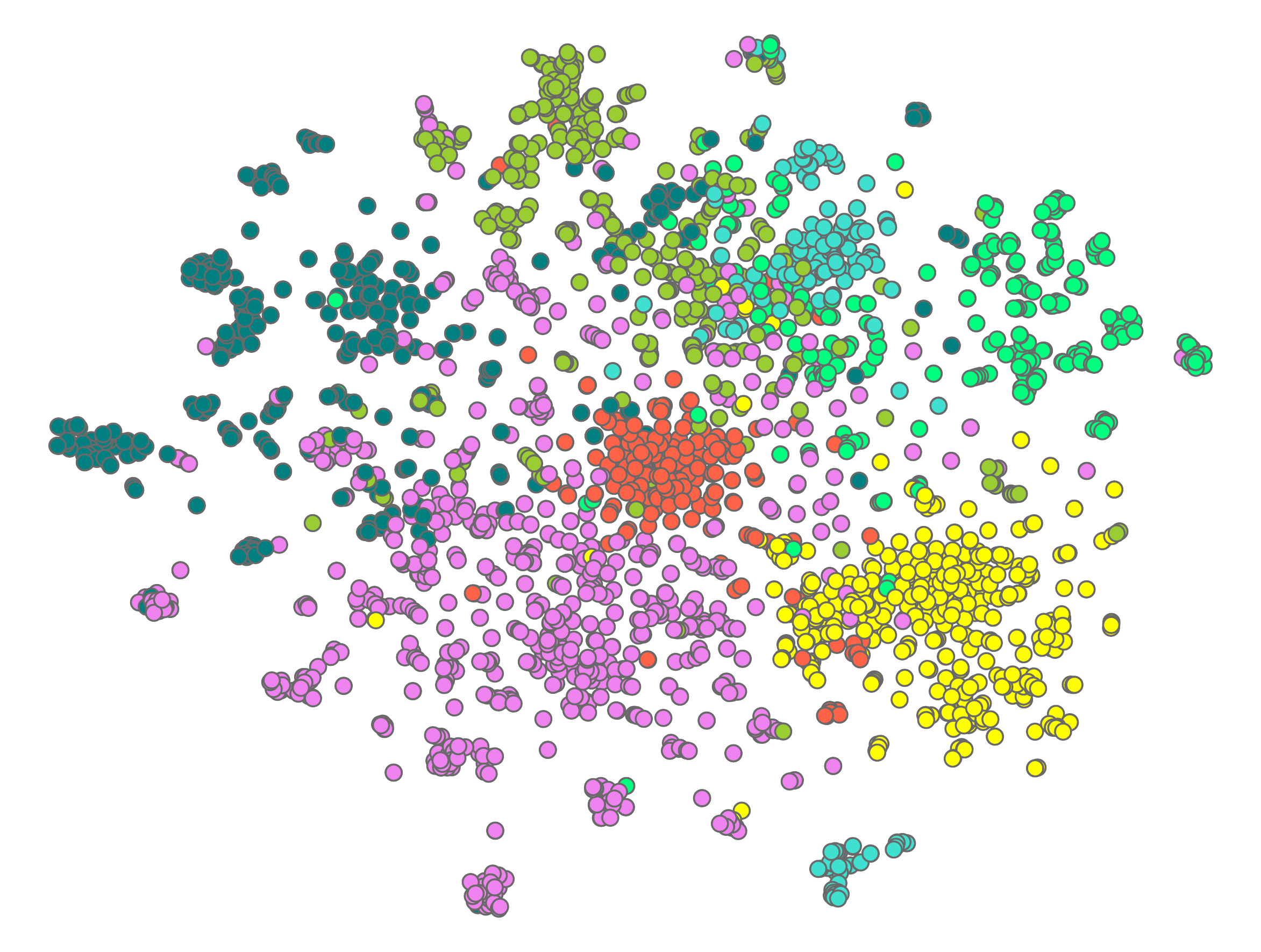}\label{fig:cora-h}}
\subfloat[FANE]{\includegraphics[width=0.2\textwidth]{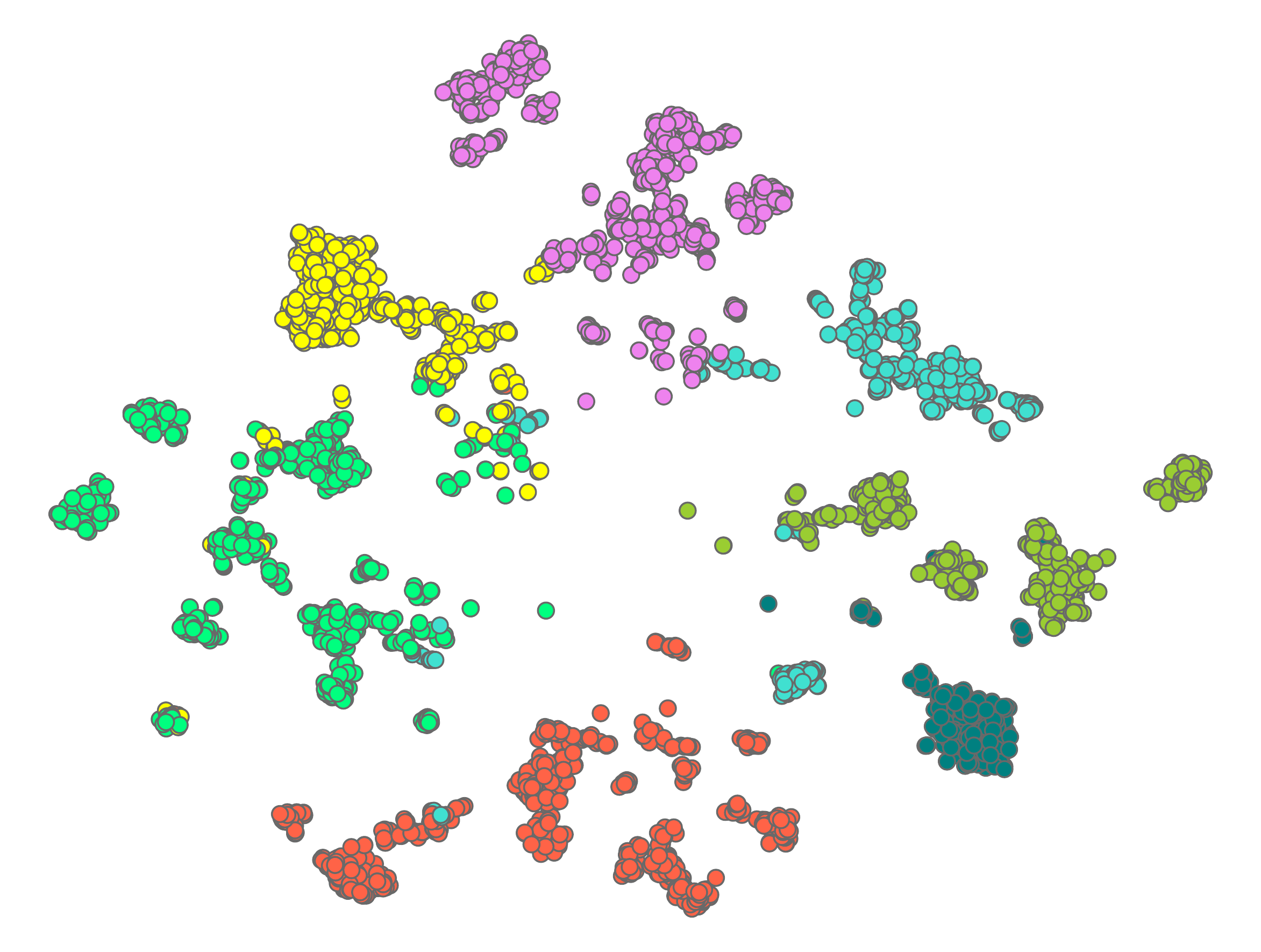}\label{fig:cora-f}}
\caption{Visualization of Cora dataset. Different colors correspond to different classes.}
\label{fig:classification-visualization}
\end{figure*}

\subsection{Classification (Quantitative Analysis)}
Network classification is one of the main application for network embedding. We test FANE on several popular network datasets with ground truth: WebKB, Cora, and Citeseer. We use Support Vector Machine (SVM) \cite{RN20150} for classification. We let the training ratio varies from 10 \% to 90\%. Other parameters are set as follows.

For node2vec, we did a grid search to find the best combination of $p$ (0.25 0.50 1.0 2.0 4.0) and $q$ (0.25 0.50 1.0 2.0 4.0) that yields the best results. The values of $p$ and $q$ are: (0.25, 2.0) in WebKB, (4.0, 2.0) in Cora, and (4.0, 0.25) in Citeseer. For struc2vec, we use the default parameters in its code. For TADW, as instructed in its code, setting textRank = 200, lambda = 0, $C$ = 5 for Cora and Citeseer. For WebKB, we tried all values of $C$ that appear in its paper and finally decide to set $C$ = 15, by which yield the best results of TADW. Similarly, for HSCA, setting textRank = 200, lambda = 0.2, mu = 0.1, $C$ = 5 for Cora and Citeseer, following the instructions given in the code. For WebKB, we tried all common values of mu and $C$ that appear in its paper and finally decided to use mu = 0.1 and $C$ = 15. Setting $d$ = 8 if the benchmark methods have the embedding dimension parameter, including FANE.

\begin{table}
\centering
 \caption{FANE parameters for datasets WebKB, Cora and Citeseer}
 \label{tab:FANE-param}
 \begin{tabular}{l|r|r|r|r}
   \toprule
   \multicolumn{1}{c|}{Datasets} & \multicolumn{1}{c|}{p} & \multicolumn{1}{c|}{q} & \multicolumn{1}{c|}{r} & \multicolumn{1}{c}{C}\\ \hline
   WebKB & 1.0 & 0.50 & 2.0 & 1.0   \\ \hline
   Cora & 3.0 & 0.15 & 2.0 & 0.1   \\ \hline
   Citeseer & 4.0 & 0.30 & 6.0 & 0.2   \\
   \bottomrule   
 \end{tabular}
\end{table}

For FANE, we set values of parameters for different datasets as shown in Table \ref{tab:FANE-param}. Note that $C$ is the penalty parameter in SVM training. Figure \ref{fig:classification} shows the Micro-F$_1$ and Macro-F$_1$ \cite{schutze2008introduction} results on the datasets. There are mainly three observations from the result:

\begin{figure*}
\centering
\subfloat[ego-Facebook]{\includegraphics[width=0.3\textwidth]{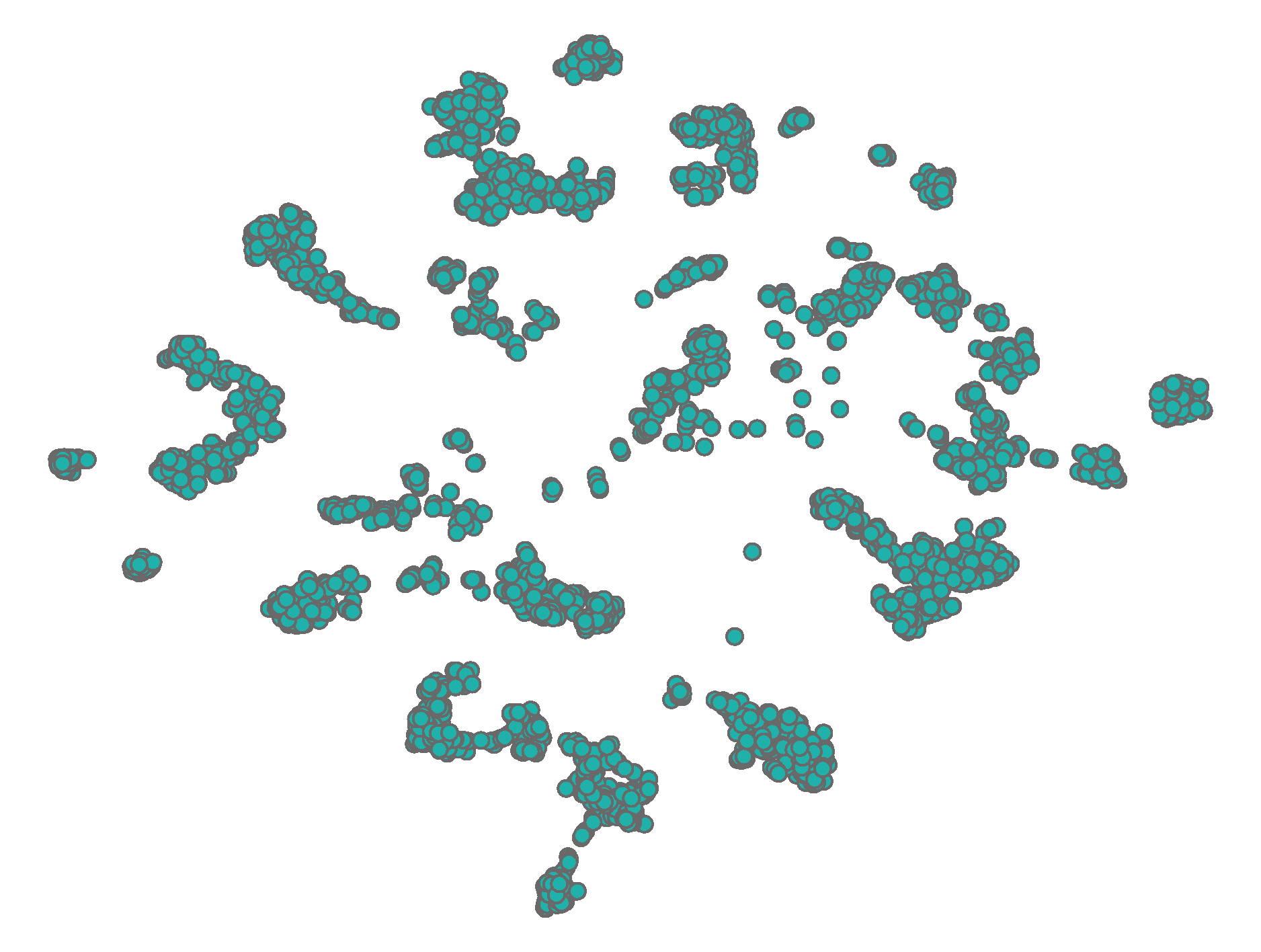}\label{fig:v-facebook}}
\subfloat[ego-Facebook attributes]{\includegraphics[width=0.3\textwidth]{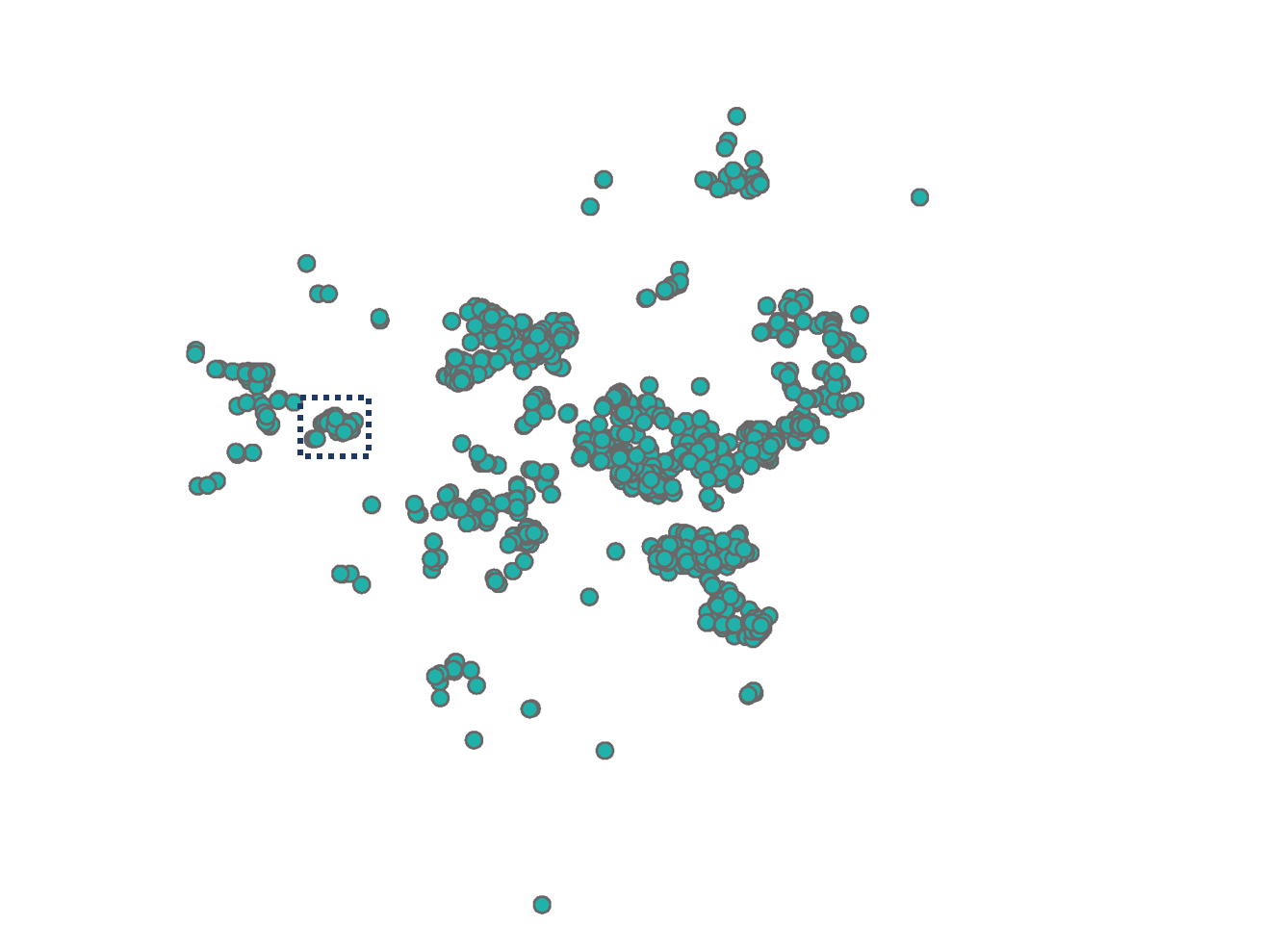}\label{fig:v-facebook-an}}
\subfloat[ego-Facebook attributes part]{\includegraphics[width=0.3\textwidth]{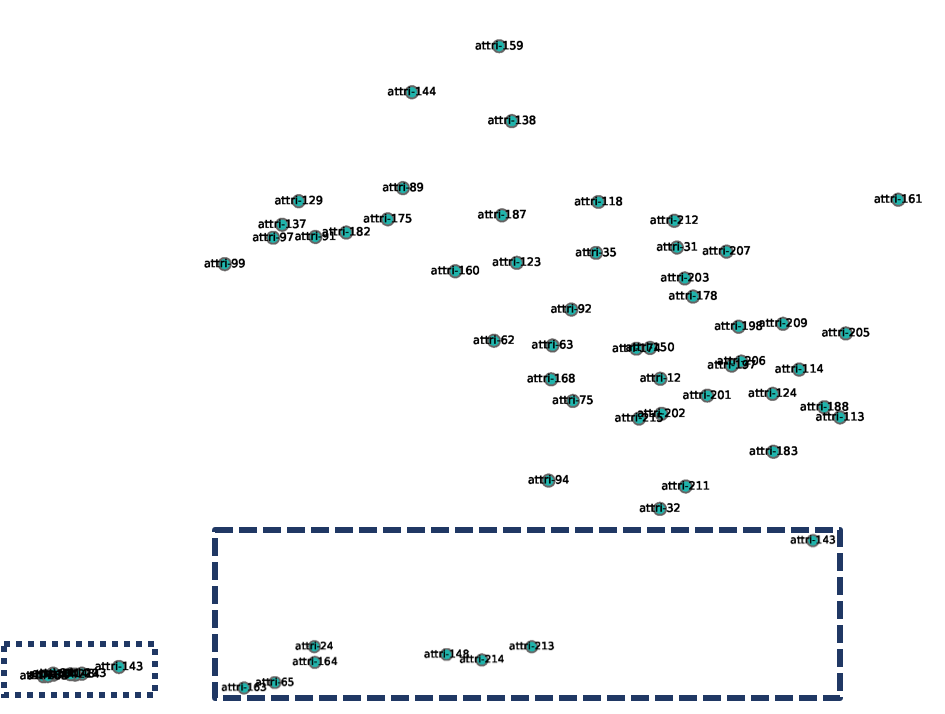}\label{fig:v-facebook-an-part}} \newline
\subfloat[ego-GPlus]{\includegraphics[width=0.3\textwidth]{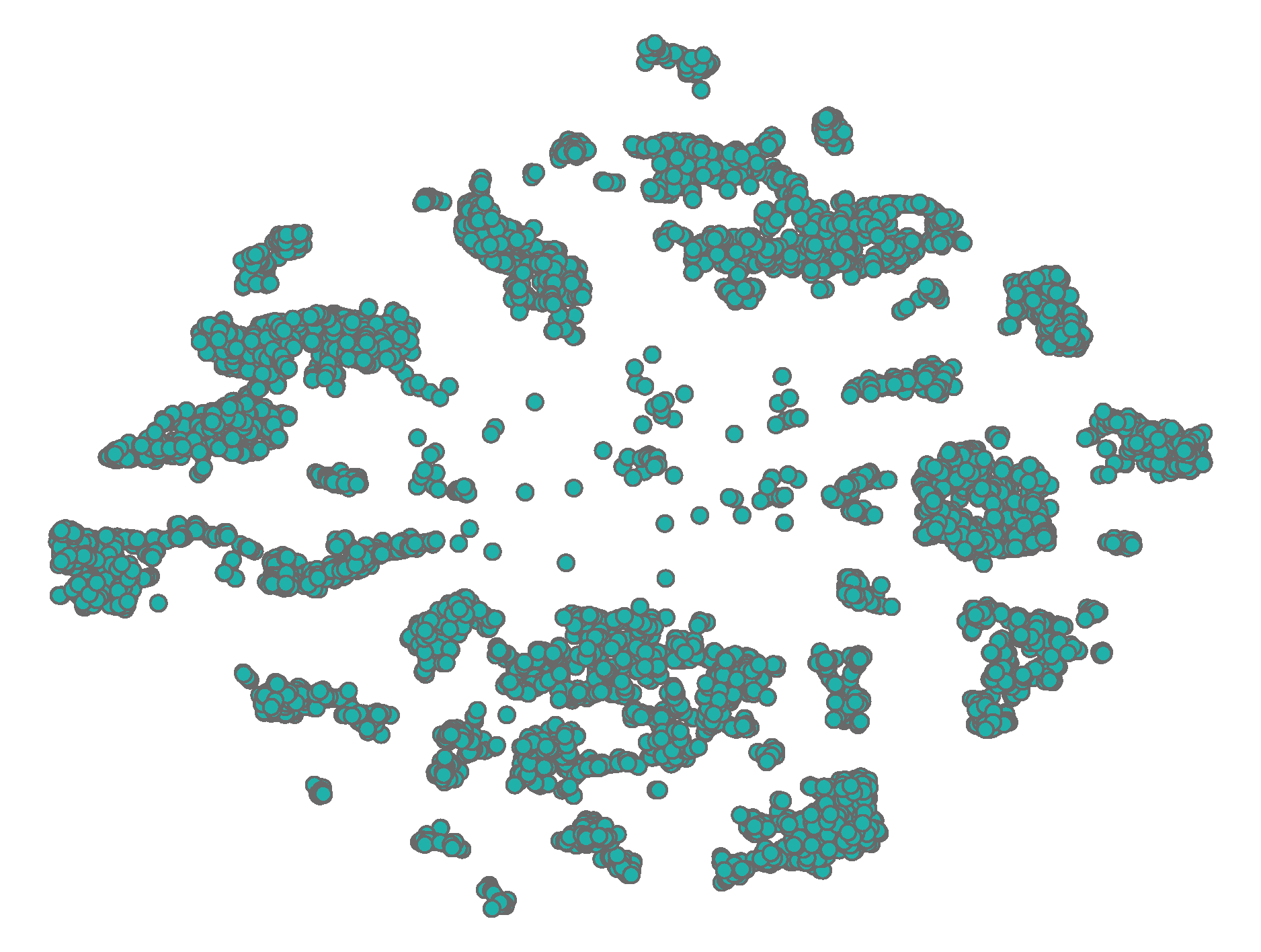}\label{fig:v-google}}
\subfloat[ego-GPlus attributes]{\includegraphics[width=0.3\textwidth]{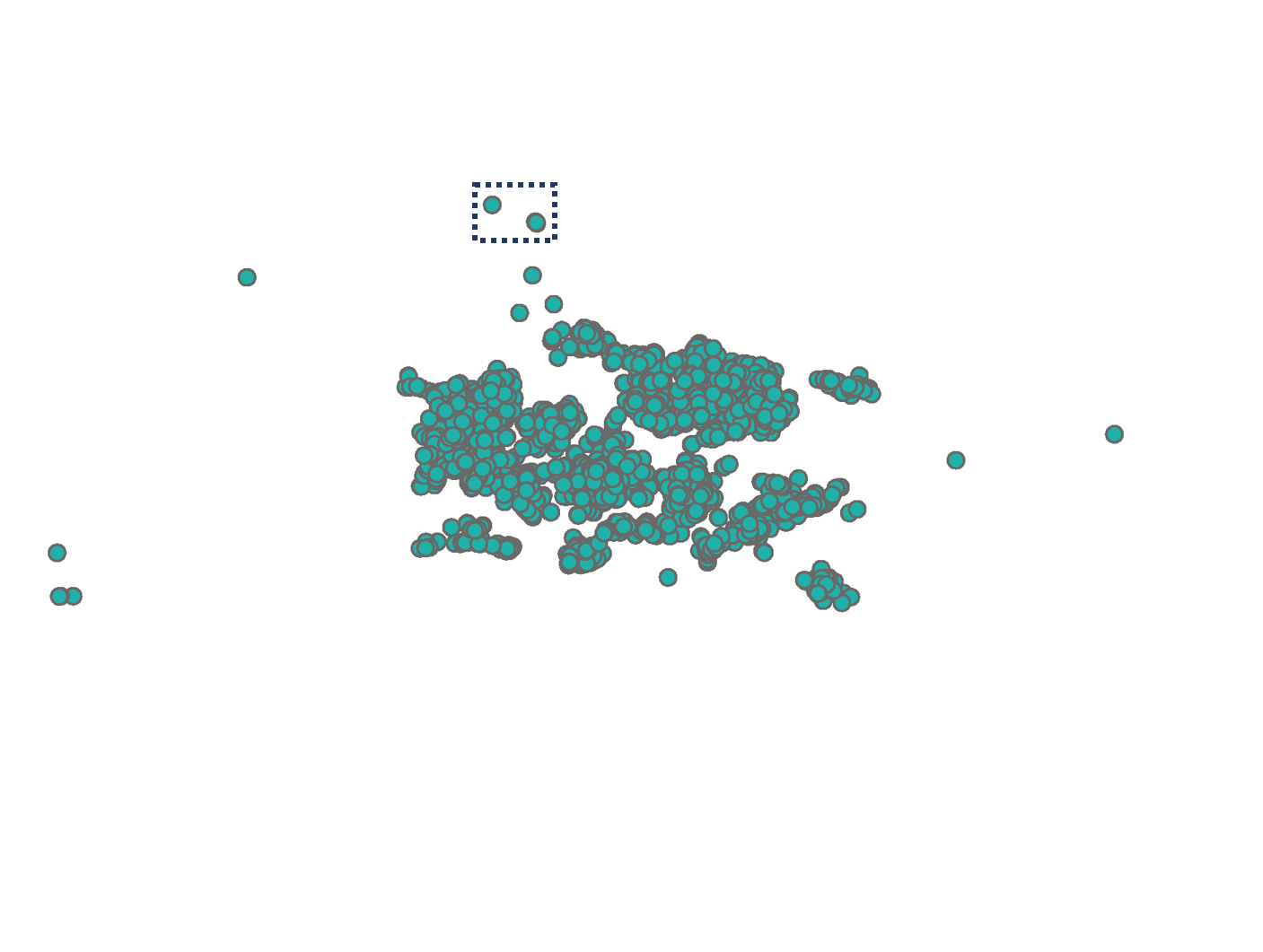}\label{fig:v-google-an}}
\subfloat[ego-GPlus attributes part]{\includegraphics[width=0.3\textwidth]{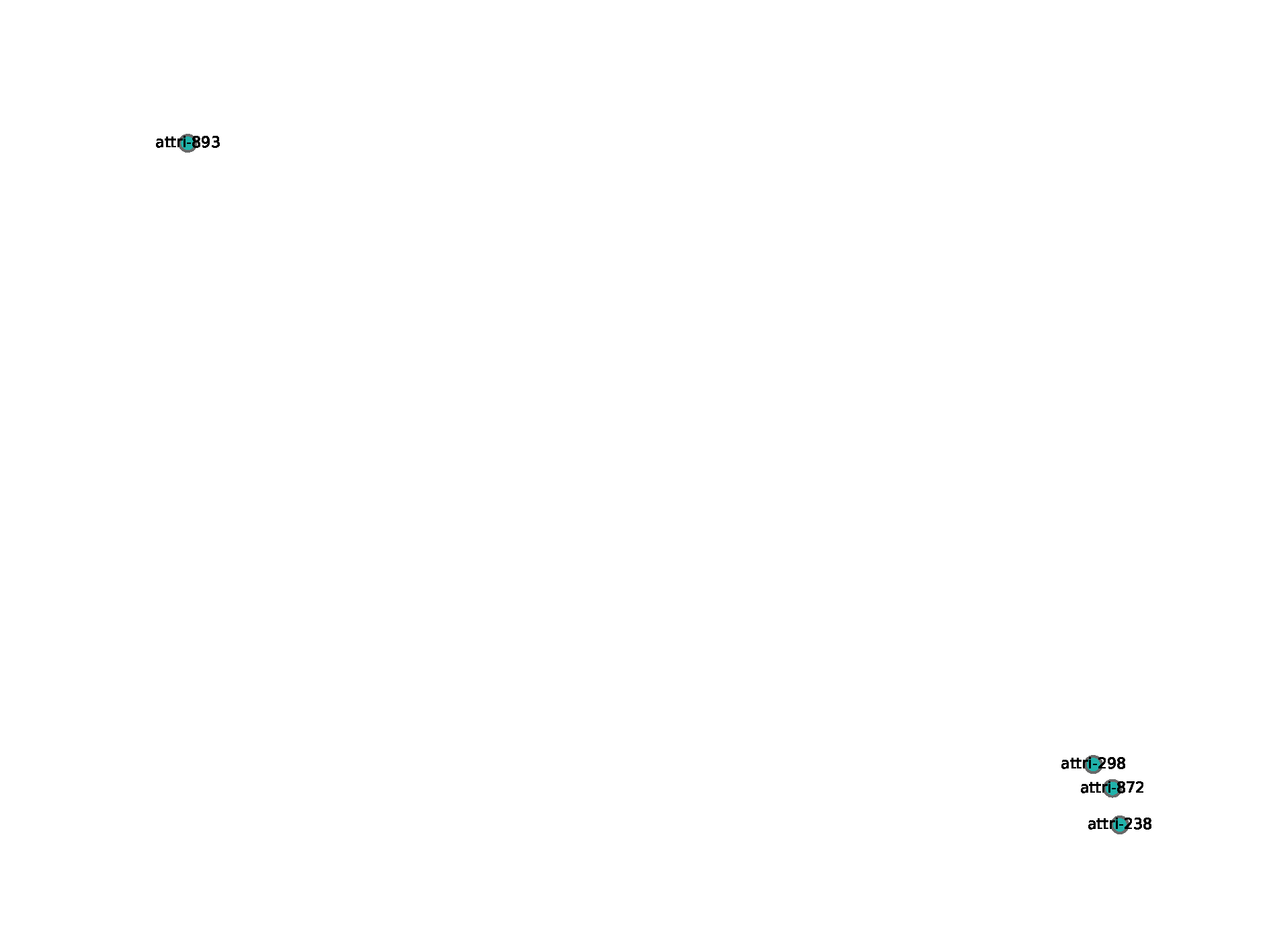}\label{fig:v-google-an-part}} \newline
\subfloat[ego-Twitter]{\includegraphics[width=0.3\textwidth]{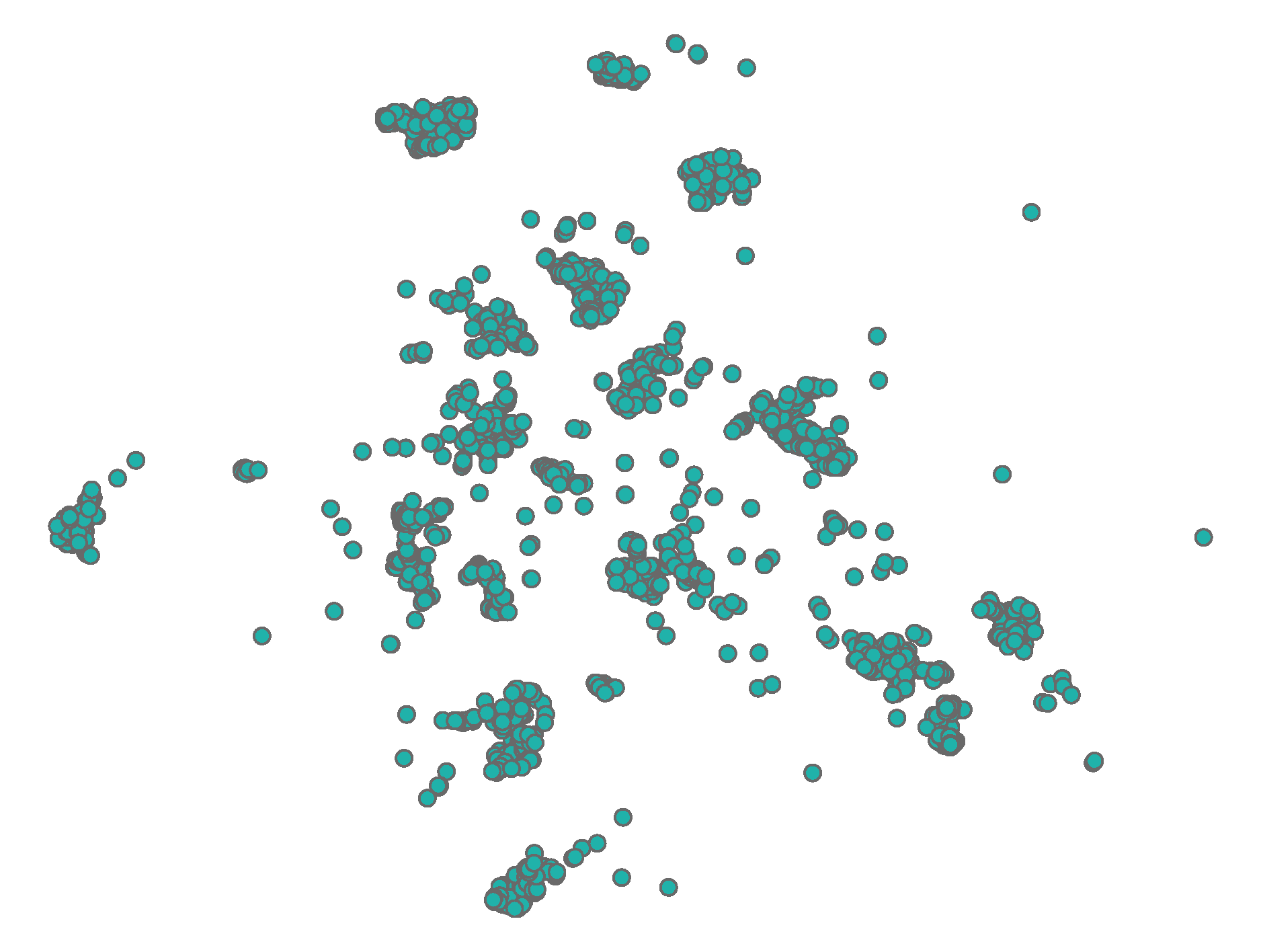}\label{fig:v-twitter}}
\subfloat[ego-Twitter attributes]{\includegraphics[width=0.3\textwidth]{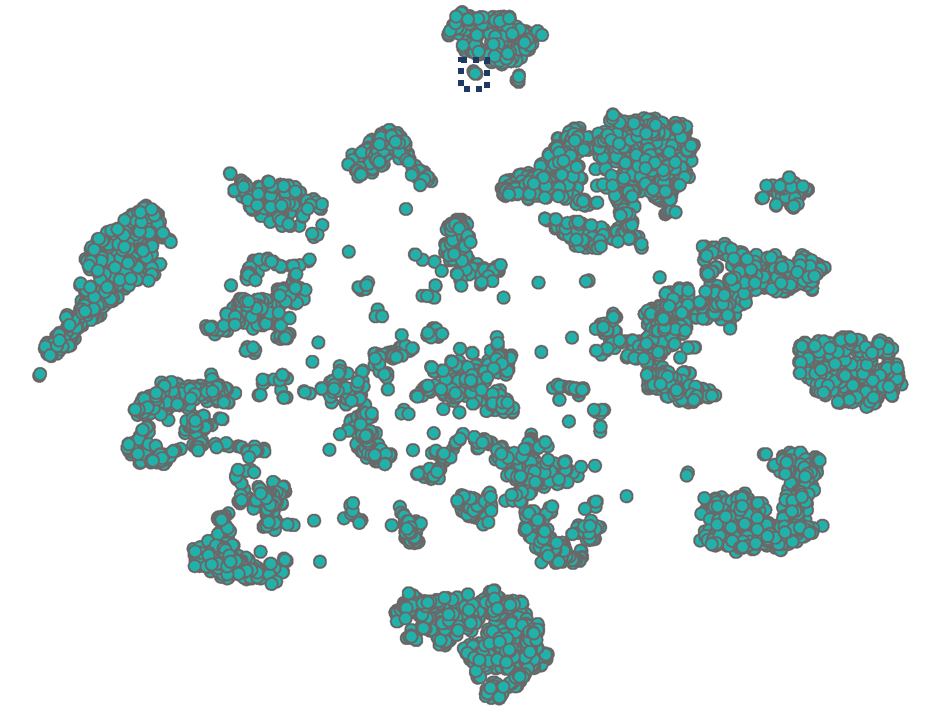}\label{fig:v-twitter-an}}
\subfloat[ego-Twitter attributes part]{\includegraphics[width=0.3\textwidth]{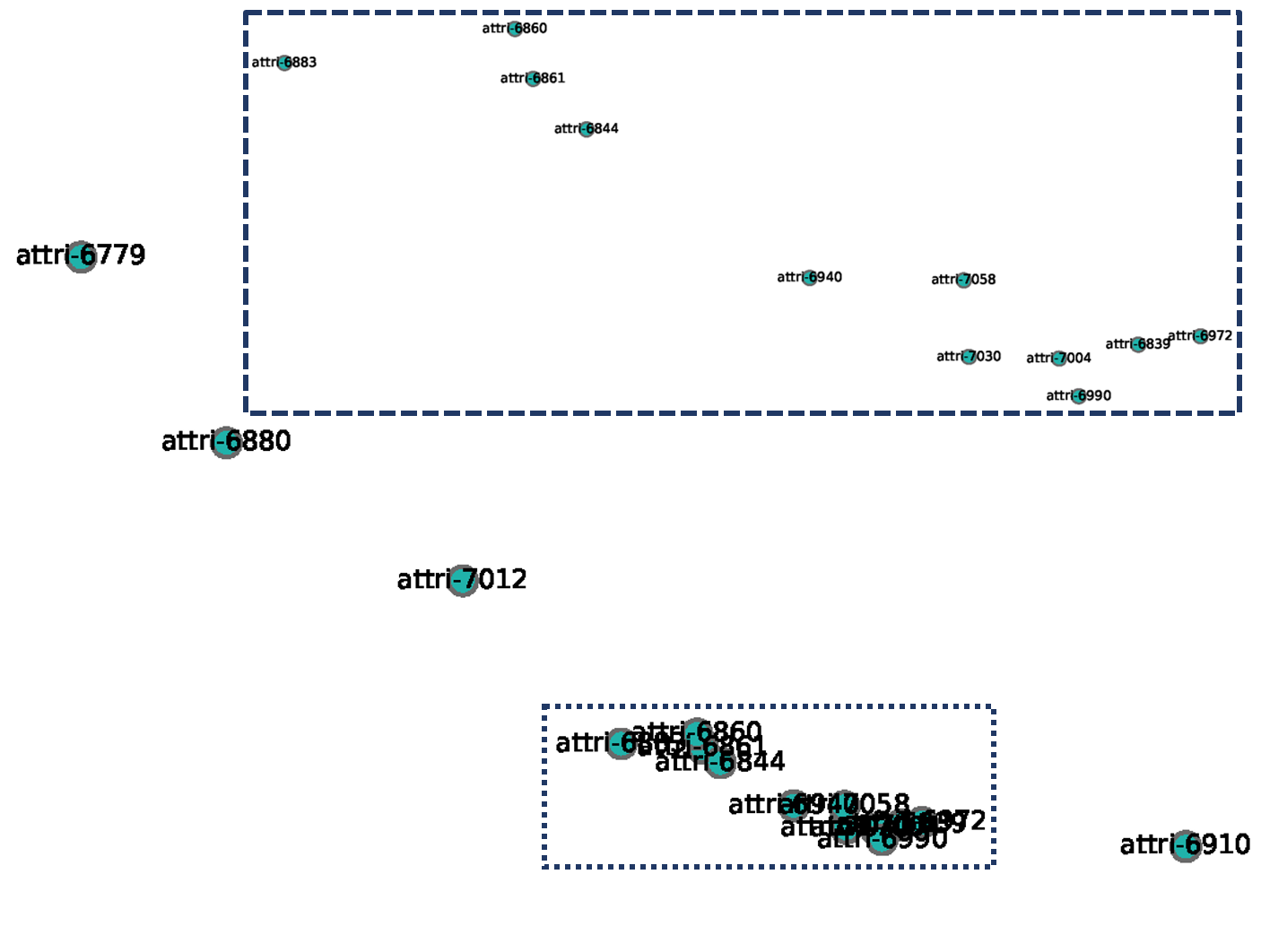}\label{fig:v-twitter-an-part}}
\caption{Visualization and Exploration of the embedding results of raw network and attributes with FANE. Left column: embedding results of ego-Facebook, ego-GPlus and ego-Twitter from top to bottom; middle column: embedding results of attributes of corresponding datasets; right column: enlarged parts of embedding results of attributes which are marked with dashed black rectangle. It is worth to notice that attribute nodes are also separate to different groups, which telling that these attributes are closely related. (Limited by the space, please magnify corresponding part to see the labels.)}
\label{fig:fgt-visualization}
\end{figure*}

\begin{enumerate}
\item FANE consistently outperforms all other benchmark methods for all datasets and for all training data. In dataset Cora, FANE even has average 10\% to 20\% better performance than that of HSCA and node2vec, which are the second highest results, under the same training ratio. It is worth mentioning that even at very low training ratio, FANE can still yields high accuracy. For example, in dataset WebKB and Cora, our method with training data equal to 10\% could match or even beat other benchmark methods with 90\% training data.
\item FANE's performance of classification significantly increase as the number of samples increases. Notice that the benchmark methods generally reach their extreme after certain number of training samples. For example, in dataset Cora, the classification accuracy of the benchmark methods start to fluctuate up and down beyond 60\% training ratio. Similarly, in dataset Citeseer, the fluctuation patterns even start at 10\% training data. On the contrary, FANE still yields a significant increase in classification accuracy as the training samples increase.
\item FANE has more significant improvement for Macro-F$_1$ in cases like WebKB and Cora. As shown in Figure \ref{fig:classification}, FANE outperforms HSCA, which is the second best results, for nearly averaged 20\%. Macro-F$_1$ treats each class equally and computes the average result of each class. Thus, a high Macro-F$_1$ result implies that FANE accurately classify nodes for their ground truths for every class.
\end{enumerate}

Figure \ref{fig:classification-visualization} shows a visual illustration of classification results. The embedding results are classified with K-Means. As we can see, the different classes are reasonably separated after classification with FANE. Furthermore, recall that our main purpose is to provide a flexible framework to realize the transition and the integration between structural-preserving and attribute-preserving. The parameter $r$ is free to be decreased or increased depending which information we want to preserve.

\subsection{Visualization (Qualitative Analysis)}
Network Visualization has been widely used in network interaction and analysis. It is a powerful tool to reveal the content of a network in a easily interpretable way by finding patterns, marking connections, and showing clustering results. Besides visual analysis of network as existing methods. our method are inherently suitable to visual analyze network attributes.

The visual analysis technique helps us design, evaluate, and explore our proposed framework. To be more specific, Figure \ref{fig:attri2vec-f} to Figure \ref{fig:adjnoun-r} help us visualize our different random walking strategies and their effects on network embedding. Thus, we can directly see how changing strategies or parameters alters the embedding results. This offers great help in the designing process. Moreover, Figure \ref{fig:comparation} and Figure \ref{fig:classification-visualization} provide another viable way to evaluate our results: we can now straightforwardly view how the embedded nodes are mixed, separated, and classified as different groups.

Figure \ref{fig:fgt-visualization} also offer some insights to be explored. For example, Figure \ref{fig:v-facebook}, Figure \ref{fig:v-google}, and Figure \ref{fig:v-twitter} demonstrate the embedding result for dataset ego-Facebook, ego-Gplus, and ego-Twitter. As we can see, nodes are grouped into different areas. Recall that FANE treat attribute as nodes during the embedding process. Therefore, we can also adopt the same visual analysis techniques to explore information about attributes in a 2D figure as shown in Figure \ref{fig:v-facebook-an}, Figure \ref{fig:v-google-an}, and Figure \ref{fig:v-twitter-an}. We find the visualized patterns for those attributes are interesting to be discussed.

For attributes in Facebook and Twitter, we find that attributes are also separated to different groups just like those embedded nodes in Figure \ref{fig:v-facebook} and \ref{fig:v-twitter}. Thus, we enlarge the small selected region in Figure \ref{fig:v-facebook-an} to see if we can uncover some relationships between those attributes. In Figure \ref{fig:v-facebook-an-part}, attribute number 163(work; end\_date), 65(education; year), 24(education; school), 164(work; start\_date), 148(work; employer), 214(education; concentration), 213(education;concentration), and 143(work;employer) \footnote{213 and 214 represent two different concentrations (majors) while 143 and 148 represent two different employers.} are closely grouped together. This result is not surprising because we know one's majors is somehow related to his/her schools and decide which type of enterprises they will enter in the future. Besides, the years they graduate from schools also affect the years they start or end their works. Even though all attributes in the Facebook dataset are anonymous, it is interesting to see that attributes are grouped based on their information.

Nevertheless, the embedded attribute pattern for Google is different to those of Facebook and Twitter: most of the attributes are grouped together with only few attributes placed far away from them. We also enlarge some of the isolated attributes to see if attributes in those isolated areas are still grouped based on their meanings. There are four attributes in that selected area in Figure \ref{fig:v-google-an-part}: 893(job\_title: music), 298(job\_title: dj), 872(job\_title: dj) \footnote{893 and 298 means two different attributes with same type of jobs}, and 238(job\_title: producer). There is an intense correlation among working as DJ, working in the field of music, and working as a producer. This example confirms that attributes are also grouped based on their meaning even in this isolated area.

\begin{figure}
\centering
\includegraphics[width=0.48\textwidth]{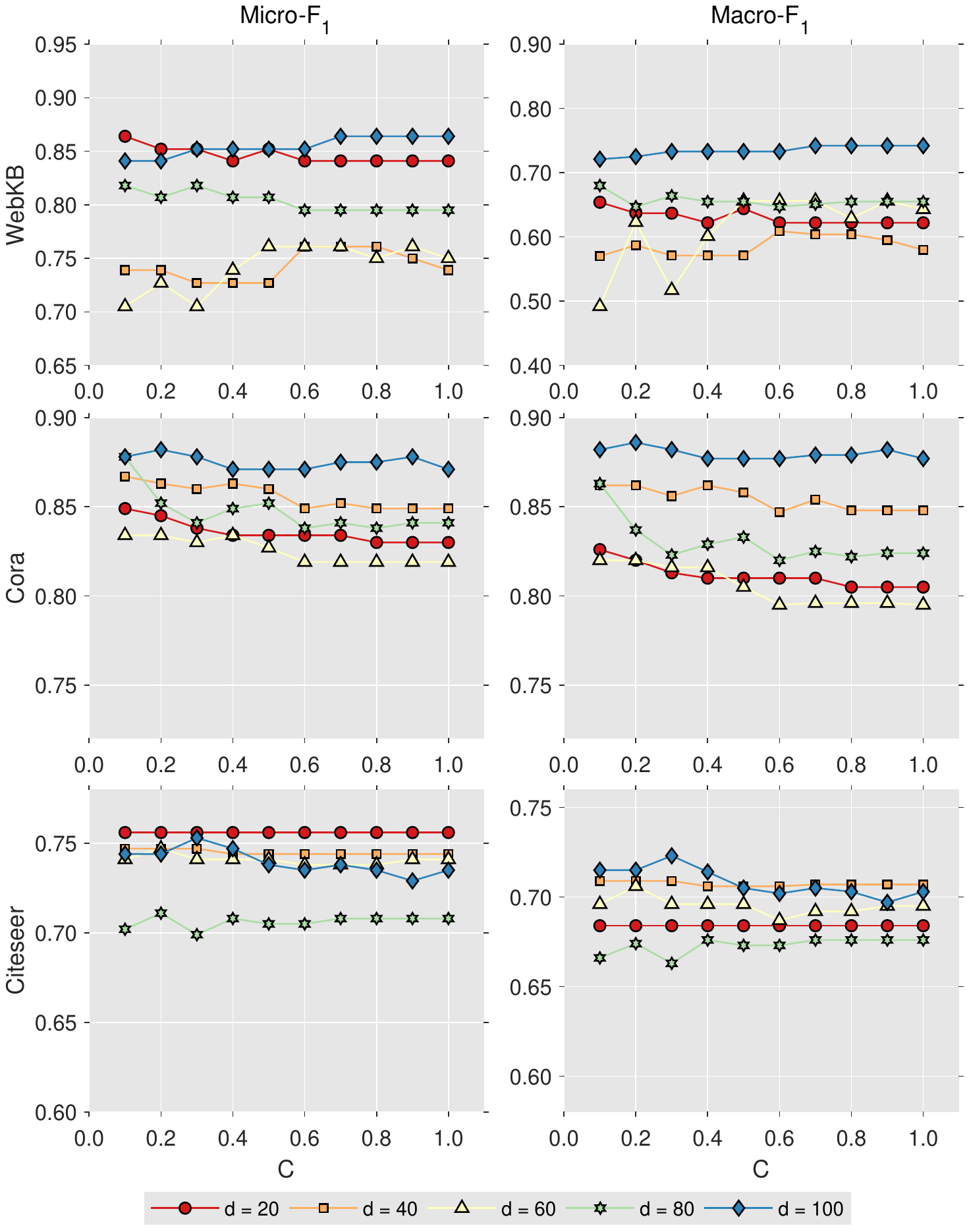}
\caption{Parameter sensitivities of embedding dimension, $d$, and penalty parameter, $C$, in FANE.}
\label{fig:param-sensitivity}
\end{figure}

We keep exploring the attributes information for Twitter. Notice that unlike those of Facebook or Google+, attributes in Twitter are actually hashtags and users. Consequently, we noticed some patterns in which hashtags and users are clustered according to their contents and their personal interests. In Figure \ref{fig:v-twitter-an-part}, all attributes are related to BBC in terms of BBC News and BBC Sports. For example, 7004(@bbcsport) is the official BBC Sport channel in Twitter, 7030(@georgeyboy) is a British famous sport broadcaster and used to work for BBC 5 live, and 6990(@bbc5live) is the official channel for BBC 5 Live. Thus, different types of attributes, personal information or user themselves, could be embedded and analyzed as long as there exits some shared information between them.

From the above examples, we proved that it is possible to uncover some nontrivial information about attributes if we also embed the attribute and visualize them. Thus, our proposed framework, FANE, is not limited in flexibly integrating attribute and structural information of the nodes; instead, it could be extended to reflect more information of attributes themselves.

\subsection{Parameter Sensitivity}
To test the sensitivities of parameter $d$ and $C$, we fixed the values of $p, q, r$, and $t$ for datasets Cora, Citeseer, and Wiki. We let the embedding dimension, $d$, vary from 20 to 100 and let the penalty parameter $C$ vary from 0.1 to 1.0. The result is shown in Figure \ref{fig:param-sensitivity}, where the x-axis represents the variations of values of $C$ and the different lines represents the variations of values of $d$.

As we can see, the testing accuracies is relatively stable with respect to the penalty parameter, $C$, in most cases when the embedding dimension is kept unchanged. However, in some cases, $C$ also heavily impacts the result. For example, in Dataset WebKB $d = 60$, testing result fluctuate over 15\% with values of $C$ changed. Moreover, the embedding dimension, $d$, has a relatively high impact on the testing accuracy in Figure \ref{fig:param-sensitivity}. Overall, our testing results are relatively stable regard to the values of $C$ while are heavily impacted by the values of $d$.

\subsection{Scalability}
\begin{figure}
\centering
\includegraphics[width=0.48\textwidth]{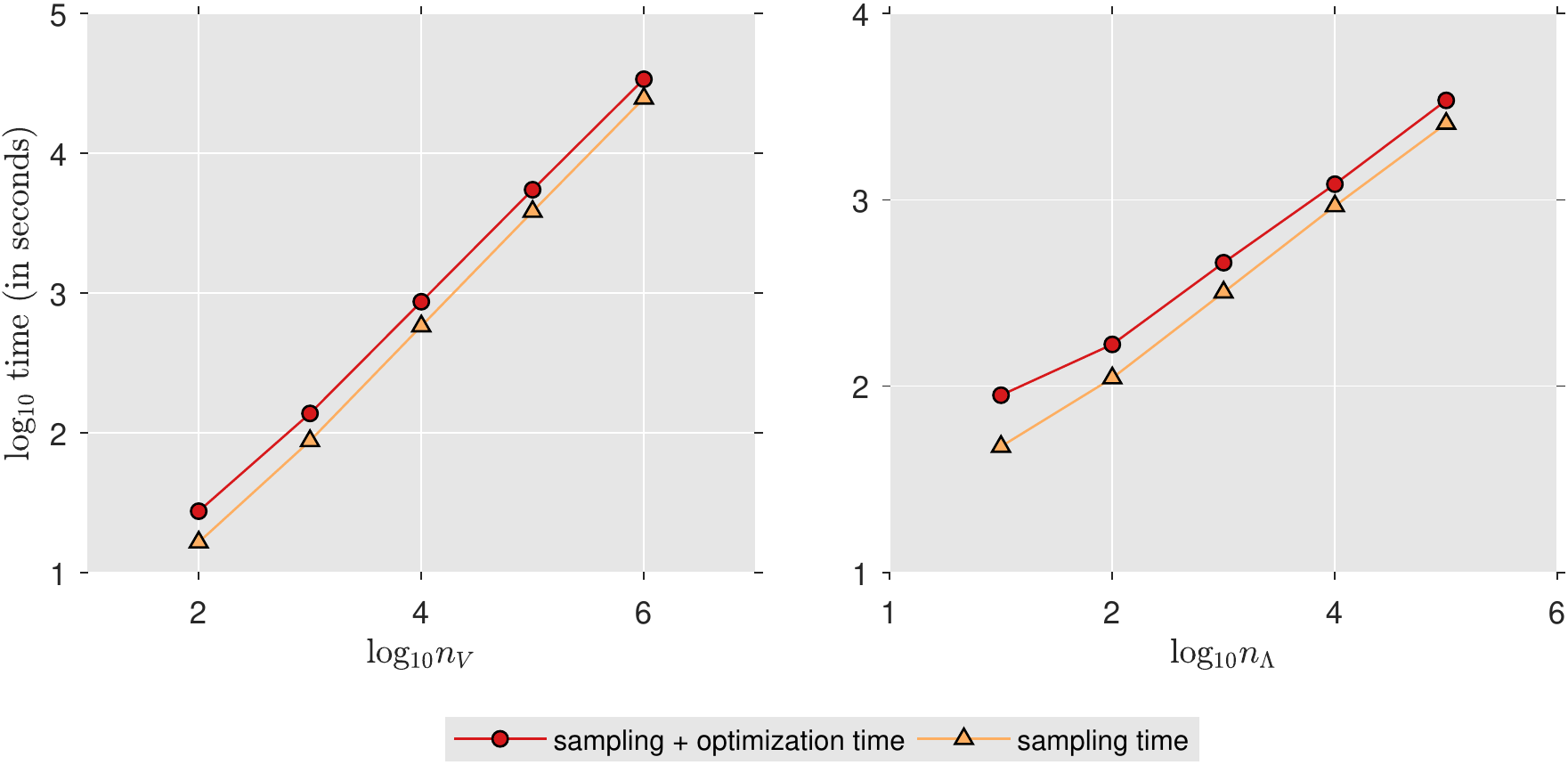}
\caption{Scalability of FANE on Erdos-Renyi network.}
\label{fig:scalability}
\end{figure}

To evaluate the scalability of FANE, we learn representations from Erdos-Renyi graphs with increasing node size ranging from $10^2$ to $10^6$ at degree of 10. As we can see, in Figure \ref{fig:scalability}, the computational time scales up linearly with increasing number of nodes. Recall that $n_V$ and $n_\Lambda$ stand for number of nodes and attributes separately. This result proves that our proposed framework is scalable with respect to number of nodes.

Since we treat attributes as nodes during our network embedding process, we also test the scalability with size of attributes per node. Figure \ref{fig:scalability} plot a network of computational time versus number of attribute nodes to fully prove the scalability of our framework. Based on Erdos-Renyi network, we fix the number of nodes to 1000 and add the number of attributes from $10$ to $10^5$ for each nodes. As we can see, the computational time also vary linearly with the number of attribute nodes. Consequently, these two experiments confirm that our proposed framework is scalable over number of nodes and number of attributes. Thus, our method could handle large scale network embedding with controllable amount of time.

\section{Conclusion} \label{sec:conclusion}
In this paper, we proposed an attributed network embedding framework which could flexibly integrate structure information and attribute information. Thus, it could learn features based on structure, attributes or both, and could provide a smooth transition between attribute-preserving and structure-preserving embedding.

Experiments confirm that our proposed method outperforms the listed STAR methods on network classification. To our best knowledge, FANE is the first method to provide flexible adjustment between attribute-preserving and structural-preserving. Under our proposed framework, we can actively intervene the embedding process to determine which type of information or which kind of integration we want. Moreover, we provide a visual analysis approach to design, optimize, and evaluate our method. This intuitive way is non-trivial in network analysis and interaction.

In this paper, we restrict our discussions on undirected attributed graphs but our method can be easily extended to process directed attributed graphs. In addition, we assume that every attribute shares same importance. However, the attribute parameter $r$ could also be extended to reflect the relative importance between different attributes. Moreover, we treat attribute as normal nodes, it is interesting to process attribute further, such as classification or clustering. It maybe useful in problems such as attribute compression, which is important in processing network with thousands attributes.

{\small
\bibliographystyle{ieee}
\bibliography{egbib}

\begin{thebibliography}{10}\itemsep=-1pt

\bibitem{RN20140}
C.~C. Aggarwal and H.~Wang.
\newblock A survey of clustering algorithms for graph data.
\newblock In {\em Managing and Mining Graph Data}, pages 275--301. Springer,
  2010.

\bibitem{RN20141}
R.~K. Ahuja, T.~L. Magnanti, and J.~B. Orlin.
\newblock {\em Network flows - theory, algorithms and applications}.
\newblock Prentice Hall, 1993.

\bibitem{RN20012}
H.~Cai, V.~W. Zheng, and K.~C. Chang.
\newblock A comprehensive survey of graph embedding: Problems, techniques, and
  applications.
\newblock {\em IEEE Transactions on Knowledge and Data Engineering},
  30(9):1616--1637, 2018.

\bibitem{RN19978}
H.~Cheng, Y.~Zhou, and J.~X. Yu.
\newblock Clustering large attributed graphs: {A} balance between structural
  and attribute similarities.
\newblock {\em {TKDD}}, 5(2):12:1--12:33, 2011.

\bibitem{Cui2018}
P.~Cui, X.~Wang, J.~Pei, and W.~Zhu.
\newblock A survey on network embedding.
\newblock {\em CoRR}, abs/1711.08752, 2017.

\bibitem{RN20145}
L.~V. Der~Maaten and G.~E. Hinton.
\newblock Visualizing data using t-sne.
\newblock {\em Journal of Machine Learning Research}, 9:2579--2605, 2008.

\bibitem{Du2018}
L.~Du, Z.~Lu, Y.~Wang, G.~Song, Y.~Wang, and W.~Chen.
\newblock Galaxy network embedding: {A} hierarchical community structure
  preserving approach.
\newblock In {\em Proceedings of the Twenty-Seventh International Joint
  Conference on Artificial Intelligence}, pages 2079--2085, 2018.

\bibitem{GaoH18b}
H.~Gao and H.~Huang.
\newblock Deep attributed network embedding.
\newblock In {\em Proceedings of the Twenty-Seventh International Joint
  Conference on Artificial Intelligence}, pages 3364--3370, 2018.

\bibitem{RN20123}
P.~Goyal and E.~Ferrara.
\newblock Graph embedding techniques, applications, and performance: {A}
  survey.
\newblock {\em Knowledge-Based Systems}, 151:78--94, 2018.

\bibitem{RN20017}
A.~Grover and J.~Leskovec.
\newblock node2vec: Scalable feature learning for networks.
\newblock In {\em Proceedings of the 22nd International Conference on Knowledge
  Discovery and Data Mining}, pages 855--864, 2016.

\bibitem{RN20104}
W.~L. Hamilton, R.~Ying, and J.~Leskovec.
\newblock Representation learning on graphs: Methods and applications.
\newblock {\em IEEE Data Engineering Bulletin}, 40(3):52--74, 2017.

\bibitem{RN20150}
M.~A. Hearst, S.~T. Dumais, E.~Osuna, J.~Platt, and B.~Scholkopf.
\newblock Support vector machines.
\newblock {\em IEEE Intelligent Systems and their Applications}, 13(4):18--28,
  July 1998.

\bibitem{Huang-etal17Accelerated}
X.~Huang, J.~Li, and X.~Hu.
\newblock Accelerated attributed network embedding.
\newblock In {\em Proceedings of the 2017 {SIAM} International Conference on
  Data Mining}, pages 633--641, 2017.

\bibitem{RN20129}
L.~Liao, X.~He, H.~Zhang, and T.~Chua.
\newblock Attributed social network embedding.
\newblock {\em CoRR}, abs/1705.04969, 2017.

\bibitem{RN20148}
Q.~Lu and L.~Getoor.
\newblock Link-based classification.
\newblock In {\em Proceedings of the 20th International Conference on Machine
  Learning}, pages 496--503, 2003.

\bibitem{schutze2008introduction}
C.~D. Manning, P.~Raghavan, and H.~Sch{\"{u}}tze.
\newblock {\em Introduction to information retrieval}.
\newblock Cambridge University Press, 2008.

\bibitem{RN20248}
P.~V. MARSDEN and N.~E. FRIEDKIN.
\newblock Network studies of social influence.
\newblock {\em Sociological Methods \& Research}, 22(1):127--151, 1993.

\bibitem{RN20118}
J.~J. McAuley and J.~Leskovec.
\newblock Learning to discover social circles in ego networks.
\newblock In {\em The Twenty-Fifth Annual Conference on Neural Information
  Processing Systems}, pages 548--556, 2012.

\bibitem{RN20249}
M.~McPherson, L.~Smith-Lovin, and J.~M. Cook.
\newblock Birds of a feather: Homophily in social networks.
\newblock {\em Annual Review of Sociology}, 27(1):415--444, 2001.

\bibitem{RN20126}
T.~Mikolov, K.~Chen, G.~Corrado, and J.~Dean.
\newblock Efficient estimation of word representations in vector space.
\newblock {\em CoRR}, abs/1301.3781, 2013.

\bibitem{RN20144}
M.~E.~J. Newman.
\newblock Finding community structure in networks using the eigenvectors of
  matrices.
\newblock {\em Physical Review E}, 74(3):036104, 2006.

\bibitem{RN20110}
B.~Perozzi, R.~Al{-}Rfou, and S.~Skiena.
\newblock Deepwalk: online learning of social representations.
\newblock In {\em Proceedings of the 20th {ACM} International Conference on
  Knowledge Discovery and Data Mining}, pages 701--710, 2014.

\bibitem{RN20143}
M.~J. Rattigan, M.~E. Maier, and D.~D. Jensen.
\newblock Graph clustering with network structure indices.
\newblock In {\em Proceedings of the 24th International Conference on Machine
  Learning}, pages 783--790, 2007.

\bibitem{RN20124}
L.~F.~R. Ribeiro, P.~H.~P. Saverese, and D.~R. Figueiredo.
\newblock \emph{struc2vec}: Learning node representations from structural
  identity.
\newblock In {\em Proceedings of the 23rd {ACM} {SIGKDD} International
  Conference on Knowledge Discovery and Data Mining}, pages 385--394, 2017.

\bibitem{RN20147}
P.~Sen, G.~Namata, M.~Bilgic, L.~Getoor, B.~Gallagher, and T.~Eliassi{-}Rad.
\newblock Collective classification in network data.
\newblock {\em {AI} Magazine}, 29(3):93--106, 2008.

\bibitem{RN20125}
J.~Tang, M.~Qu, M.~Wang, M.~Zhang, J.~Yan, and Q.~Mei.
\newblock {LINE:} large-scale information network embedding.
\newblock In {\em Proceedings of the 24th International Conference on World
  Wide Web}, pages 1067--1077, 2015.

\bibitem{RN20142}
L.~Tao and Y.~Zhao.
\newblock Multi-way graph partition by stochastic probe.
\newblock {\em Computers \& Operations Research}, 20(3):321--347, 1993.

\bibitem{RN20134}
C.~Yang, Z.~Liu, D.~Zhao, M.~Sun, and E.~Y. Chang.
\newblock Network representation learning with rich text information.
\newblock In {\em Proceedings of the Twenty-Fourth International Joint
  Conference on Artificial Intelligence}, pages 2111--2117, 2015.

\bibitem{RN20250}
Z.~Yang, J.~Guo, K.~Cai, J.~Tang, J.~Li, L.~Zhang, and Z.~Su.
\newblock Understanding retweeting behaviors in social networks.
\newblock In {\em Proceedings of the 19th {ACM} Conference on Information and
  Knowledge Management}, pages 1633--1636, 2010.

\bibitem{RN20136}
D.~Zhang, J.~Yin, X.~Zhu, and C.~Zhang.
\newblock Homophily, structure, and content augmented network representation
  learning.
\newblock In {\em {IEEE} 16th International Conference on Data Mining}, pages
  609--618, 2016.

\bibitem{RN20015}
D.~Zhang, J.~Yin, X.~Zhu, and C.~Zhang.
\newblock Network representation learning: {A} survey.
\newblock {\em CoRR}, abs/1801.05852, 2018.

\bibitem{RN20108}
D.~Zhou, S.~Niu, and S.~Chen.
\newblock Efficient graph computation for node2vec.
\newblock {\em CoRR}, abs/1805.00280, 2018.

\bibitem{RN20115}
Y.~Zhou, H.~Cheng, and J.~X. Yu.
\newblock Graph clustering based on structural/attribute similarities.
\newblock {\em {PVLDB}}, 2(1):718--729, 2009.

\bibitem{Zhu2018}
D.~Zhu, P.~Cui, D.~Wang, and W.~Zhu.
\newblock Deep variational network embedding in wasserstein space.
\newblock In {\em Proceedings of the 24th {ACM} {SIGKDD} International
  Conference on Knowledge Discovery {\&} Data Mining}, pages 2827--2836, 2018.

\end{thebibliography}
}

\end{document}